\RequirePackage{fix-cm}
\documentclass[smallextended]{svjour3}       % onecolumn (second format)
\usepackage{natbib}

\smartqed  % flush right qed marks, e.g. at end of proof
\usepackage{graphicx}
\journalname{Empirical Software Engineering}
\usepackage{amsmath,amssymb,amsfonts}
\usepackage{algorithm2e}[ruled]
\usepackage{textcomp}
\usepackage{rotating}
\usepackage{multirow}
\usepackage{booktabs}
\usepackage{tabularx}
\usepackage{mdframed}
\usepackage{lscape}
\usepackage{longtable}
\usepackage{tablefootnote}

\usepackage{listings}
\usepackage{xcolor}
\usepackage{framed}
\usepackage{caption}
\usepackage{subcaption}
\usepackage{dsfont}
\usepackage{adjustbox}

\usepackage{tikz}
\usetikzlibrary{shapes,arrows,positioning,decorations.pathreplacing}

\newcommand\YAMLcolonstyle{\color{red}\mdseries}
\newcommand\YAMLkeystyle{\color{black}\bfseries}
\newcommand\YAMLvaluestyle{\color{blue}\mdseries}

\makeatletter

\newcommand\language@yaml{yaml}

\expandafter\expandafter\expandafter\lstdefinelanguage
\expandafter{\language@yaml}
{
  captionpos=b,
  frame=single,
  keywords={true,false,null,y,n},
  keywordstyle=\color{darkgray}\bfseries,
  basicstyle=\YAMLkeystyle,                                 % assuming a key comes first
  sensitive=false,
  comment=[l]{\#},
  morecomment=[s]{/*}{*/},
  commentstyle=\color{purple}\ttfamily,
  stringstyle=\YAMLvaluestyle\ttfamily,
  moredelim=[l][\color{orange}]{\&},
  moredelim=[l][\color{magenta}]{*},
  moredelim=**[il][\YAMLcolonstyle{:}\YAMLvaluestyle]{:},   % switch to value style at :
  morestring=[b]',
  morestring=[b]",
  literate =    {---}{{\ProcessThreeDashes}}3
                {>}{{\textcolor{red}\textgreater}}1     
                {|}{{\textcolor{red}\textbar}}1 
                {\ -\ }{{\mdseries\ -\ }}3,
}

\usepackage{hyperref}

\begin{document}

\newcommand{\atoml}{\texttt{atoml}}
\newcommand{\rev}[1]{{#1}}

\newcommand\setrow[1]{\gdef\rowmac{#1}#1\ignorespaces}
\newcommand\clearrow{\global\let\rowmac\relax}
\clearrow

\title{Differential testing for machine learning: an analysis for classification algorithms beyond deep learning}
\titlerunning{Differential testing for machine learning}

\author{
Steffen Herbold \and
Steffen Tunkel}

\institute{
Steffen Herbold\\Institute of Software and System Engineering, TU Clausthal, Germany\\
\email{steffen.herbold@tu-clausthal.de}
\vspace{5pt}\\
Steffen Tunkel\\Institute of Computer Science, University of Goettingen, Germany\\
\email{steffen.tunkel@stud.uni-goettingen.de}
}

\date{Received: date / Accepted: date}

\maketitle

\begin{abstract}
\textit{Context:}
Differential testing is a useful approach that uses different implementations of the same algorithms and compares the results for software testing. In recent years, this approach was successfully used for test campaigns of deep learning frameworks.  

\noindent
\textit{Objective:}
There is little knowledge on the application of differential testing beyond deep learning. Within this article, we want to close this gap for classification algorithms.

\noindent
\textit{Method:}
We conduct a case study using Scikit-learn, Weka, Spark MLlib, and Caret in which we identify the potential of differential testing by considering which algorithms are available in multiple frameworks, the feasibility by identifying pairs of algorithms that should exhibit the same behavior, and the effectiveness by executing tests for the identified pairs and analyzing the deviations. 

\noindent
\textit{Results:}
While we found a large potential for popular algorithms, the feasibility seems limited because often it is not possible to determine configurations that are the same in other frameworks. The execution of the feasible tests revealed that there is a large amount of deviations for the scores and classes. Only a lenient approach based on statistical significance of classes does not lead to a huge amount of test failures.

\noindent
\textit{Conclusions:}
The potential of differential testing beyond deep learning seems limited for research into the quality of machine learning libraries. Practitioners may still use the approach if they have deep knowledge about implementations, especially if a coarse oracle that only considers significant differences of classes is sufficient. 

\keywords{machine learning \and software testing \and differential testing}
\end{abstract}

\section{Introduction}
\label{sec:introduction}

With differential testing, multiple implementations of the same algorithm are executed with the same data. Inconsistencies between the results are indicators for bugs. In recent years, this idea was adopted to define test oracles for deep learning \citep[e.g.,][]{Pham2019, Wang2020, Guo2020, Asyrofi2020}. This works well, because there are multiple versatile frameworks such as PyTorch~\citep{NEURIPS2019_9015} and TensorFlow~\citep{Abadi2016} that allow the definition of exactly the same neural network structures and training procedures. Beyond deep learning, this idea was also successfully applied to linear regression~\citep{McCullough2019}, to cases where the optimal solutions were known as oracle beforehand. Due to the importance of linear regression as a basic technique, there are many powerful implementations of this, which enabled the definition of differential testing. Moreover, the known optimal solution served as an additional test oracle. \rev{However}, even though differential testing was successful for deep learning and linear regression, \rev{it is} not obvious that this should also be the case for other machine learning algorithms. For example, while the general concept of random forests~\cite{Breiman2001} is well-defined,\footnote{bootstrap sampling of instances, subsets of features for decision} other aspects depend on the implementation, e.g., which decision tree algorithm is used and how the sub-sampling can be configured. \rev{Whether we can expect random forest implementations to behave the same depends on how developers navigate these options, e.g., if they implement the same variants in a hard-coded way or if they expose configuration options through hyperparameters.}  Due to this, it is unclear if and how different implementations of the same algorithms can be directly compared to each other. \rev{With deep learning, this problem does not exist: network structures, training procedures, loss functions, and optimization algorithms are all configured by the user of the framework through the API.} Moreover, while some deviations are always expected with randomized algorithms like the training algorithms for \rev{(deep)} neural networks, other algorithms are deterministic and should lead to exactly the same results, which was also not yet considered. 

Thus, while we know that differential testing can be useful to define pseudo oracles~\citep{Davis1981} for the quality assurance of algorithms, we lack knowledge for other types of machine learning tasks beyond deep learning and linear regression. Within our work, we close this gap for classification algorithms and investigate the following research question. 

\begin{description}
    \item[\textbf{Research Question:}] What is the \textit{potential}, \textit{feasibility}, and \textit{effectiveness} of differential testing for classification algorithms beyond deep learning, i.e., for techniques like decision trees, random forests, or $k$-nearest neighbor classification. 
\end{description}

We investigate our research question within an exploratory case study of four machine learning frameworks: Scikit-learn~\citep{Pedregosa2011}, Weka~\citep{Frank2016}, Spark MLlib\citep{meng2016mllib}, and Caret~\citep{Kuhn2018}. We use a methodology with three phases within our study. First, we identify for which algorithms we find multiple implementations. Through this, we evaluate the \textit{potential}, i.e., for how many algorithms differential testing could possibly work, due to the availability of multiple implementations in large frameworks. Second, we compare the algorithms in detail to understand if it is possible to configure them in the same way using their public interfaces through the definition of appropriate hyperparameters. Through this, we evaluate the \textit{feasibility}, i.e., if we can actually exploit the potential because implementations are sufficiently similar to serve as pseudo oracles and identify inconsistencies. Third, we implement differential testing for the algorithms for which we are able to find identical configurations. Through this, we evaluate the \textit{effectiveness} of differential testing, i.e., the capability to find inconsistencies with the goal to reveal potential bugs. 

Through this study, we contribute the following to the state-of-the-art.
\begin{itemize}
    \item We found that while there is a large potential for differential testing, due to different implementations of the same algorithms, it is difficult to harness the potential and find feasible combinations of algorithms that should behave the same. This is due to the diversity in hyperparameters between the implementations. 
    \item We observe many deviations between implementations that should yield similar results. The number of deviations indicates that there is a large amount of noise in the results, which makes it impossible for us to pick up a reliable signal to identify possible bugs. 
    \item Nevertheless, we also found that for experts that want to test a specific algorithm, an approach with a lenient test oracle that only considers if the classifications are significantly different, could be useful. 
\end{itemize}

The remainder of this article is structured as \rev{follows}. We define our terminology and notations in Section~\ref{sec:terminology} followed by a discussion of the related work in Section~\ref{sec:related-work}. Then, we present our case study in Section~\ref{sec:case-study}, including our subjects, methodology, and results for each of the three phases. We discuss the implications of our results in Section~\ref{sec:discussion}, followed by the threats to validity in Section~\ref{sec:threats} before we conclude in Section~\ref{sec:conclusion}. 

\section{Terminology and Notations}
\label{sec:terminology}

Within this article, we focus on \emph{binary classification}.  Formally, we have instances $x_{i} = (x_{i,1}, ..., x_{i,m}) \in \mathcal{F} \subseteq \mathbb{R}^m$ with labels $y_i \in \{0,1\}$ for $i=1, ..., n$. We say that $\mathcal{F}$ is the feature space. The binary set $\{0,1\}$ represents the classes and we note that the classes are considered as categories and can be replaced by any other binary set, e.g., $\{false, true\}$ or \{-1, +1\}. A classification model tries to find a function $f: \mathcal{F} \to \{0,1\}$ such that $f(x_i) \approx y_i$. Often, this is done by estimating scores for both classes $c \in \{0,1\}$ as $f_{score}^c: \mathcal{F} \to \mathbb{R}$ such that $f(x_i) = \arg\max_{c \in \{0,1\}} f_{score}^c(x_i)$. We note that the scores are often probabilities of classes and the class assignment can also be optimized using different thresholds for scores. However, \rev{neither the question if the scores represent probabilities nor the optimization of thresholds is} relevant for \rev{to} work and not further discussed. 

Based on the above definitions, a classification algorithm $A$ is an algorithm that takes as input training instances $(X^{train}, Y^{train}) = (x^{train}_i, y^{train}_i), i=1, ..., n^{train}$ and outputs functions $f, f_{score} = A(X^{train}, Y^{train})$. We further define $X^{test}, Y^{test} = (x^{test}_i, y^{test}), i=1, ..., n^{test}$ as test data. \rev{When we discuss the comparison of two algorithms, we refer to them as $A^1$ and $A^2$ with the respective functions $f^1, f^1_{score}, f^2$, and $f^2_{score}$ as result of the training. Moreover, we use the notation $\mathds{1}_{condition}$ for the indicator function, which is one when the condition is fulfilled and zero otherwise.}

We note that we use the terms false negative and false positive not with respect to the classification of the algorithms, as is common in literature about machine learning. Instead, we use these terms in their common meaning in the software testing literature: a false positive is a software test that fails, even though there is no wrong behavior and, vice versa, a false negative is a software test that misses wrong behavior it should detect. 

\section{Related Work}
\label{sec:related-work}

We restrict our discussion of related work to other studies that utilize multiple implementations of the same algorithms for the definition of software tests. There are several recent surveys on software engineering and software testing for machine learning, to which we refer readers for a general overview~\citep{Zhang2020, Braiek2020, Giray2021, martinezfernandez2021software}. \rev{This includes other approaches for addressing the oracle problem for machine learning, like metamorphic testing (i.e., systematic changes to input data such that the expected changes in output are known, e.g., \cite{Murphy2008, Xie2011, Ding2017}), using crashes as oracle~\citep[e.g.][]{Herbold2022}, defining numeric properties that can be checked~\citep[e.g.][]{Karpathy2018} or trying to map the performance of predictions to requirements to derive tests~\citep[e.g.][]{Barash2019}.} Further, there is also a different type of differential testing approach that is used for the testing of the robustness of models \citep[e.g.,][]{Pei2019, Guo2021}, i.e., similar deep learning models or inputs were used as pseudo oracles. However, such tests of specific models are not within the scope of our work, which considers the testing of the underlying libraries that implement machine learning algorithms and not the testing of learned models. 

One method to solve the test oracle problem of machine learning~\citep[e.g.][]{Murphy2007, Groce2014, marijan2020software} is to implement a differential testing between multiple implementations of the same algorithm~\cite{Murphy2007} and use the different implementations as pseudo-oracles~\citep{Davis1981}. While this idea is not new for machine learning and was already shown to work for the MartiRank algorithm~\citep{Gross2006} by \cite{Murphy2007}, there was only little follow up work in this direction until recently.

Most notably, this idea gained traction for the testing of deep learning frameworks. \cite{Pham2019} developed the tool CRADLE which used, e.g., TensorFlow~\citep{Abadi2016} and Theano~\citep{Theano2016} as backends. The comparison of results of outputs between different implementations was effective at finding errors, even though only few different deep learning models were used. Following this study, there were many papers who build upon this work, e.g., LEMON by \cite{Wang2020} that proposes a mutation approach to cover more different network architectures during the testing, AUDEE by \cite{Guo2020} that uses a genetic algorithm to generate effective tests, and CrossASR by \cite{Asyrofi2020} which applies a similar concept for the testing of automated speech recognition models. Beyond deep learning, the idea was also used by \cite{McCullough2019} for the comparison of different implementations of linear regression algorithms and shown to be effective. However, we note that while \cite{McCullough2019} compared different implementations to each other, they use analytic solutions as ground truth, i.e., they did not really use the implementations themselves as pseudo oracles. 

In comparison to prior work, our focus is not on the differential testing of a single algorithm~\citep{Murphy2007, McCullough2019} or deep learning.  Instead, we consider this question broadly for classification algorithms. With the exception of multilayer perceptrons, we exclude deep learning, which can also be used for classification, as these are already studied.\footnote{Multilayer perceptrons are commonly found in general purpose machine learning libraries but do not provide the same flexibility as deep learning frameworks.} We further note that neither \cite{Murphy2007} nor \cite{McCullough2019} studied classification algorithms, but rather a ranking algorithm and a regression algorithm. To the best of our knowledge, this is, therefore, the first work that considers differential testing for non-deep learning classification algorithms.

\section{Case Study}
\label{sec:case-study}

Within this section, we discuss the main contribution of our work: a case study about four machine learning libraries that explores the usefulness of differential testing for classification algorithms. In the following, we first discuss the subjects of our study followed by a description of our methodology, including the variables we measure and the research methods we use. Then, we proceed with the presentation of the results for each phase of our case study. We provide our implementation of the tests and additional details through our replication kit online.\footnote{\url{https://github.com/sherbold/replication-kit-2022-non-dl-diff-test} - We will create a long-term archive on Zenodo of the replication kit which we reference here in case of acceptance.}

\subsection{Subjects}

We used purposive sampling~\citep{Patton2014} for our case study based on four criteria. First, we wanted to use subjects implemented in different programming languages. One important use case of differential testing is to validate the behavior of new implementations of algorithms, which includes the implementation in a new programming language. Second, all subjects should be implemented by, to the best of our knowledge, independent project teams. Overlap between teams could have an adverse affect on our study of differential testing, because our results could be affected by the same developers working on multiple implementations. \rev{There is a non-negligible likelihood that such a developer would make similar design and implementation choices, which could lead to an overestimation of similarities between independent implementations.} Third, subjects should cover multiple algorithms. If our subjects are too small, this would negatively affect our research as the likelihood of finding the same algorithms multiple times would decrease. Fourth, the libraries should not focus on deep learning. Since our goal is to provide a perspective beyond deep learning, we specifically exclude libraries whose main use case is deep learning, even if the methods provided by these libraries could also be used to implement, e.g., a linear logistic regression model. \rev{Based on these criteria, we selected four such libraries as shown in Table~\ref{tbl:-frameworks}.} 

%\begin{itemize}
%\item Weka~\citep{Frank2016} is a popular machine learning library for Java which had the first official release in 1999. The library is very mature and has been used by researchers and practitioners for decades and is, e.g., part of the Pentaho business intelligence software.\footnote{https://www.hitachivantara.com/en-us/products/data-management-analytics/pentaho-platform.html} 
%\item Scikit-learn~\citep{Pedregosa2011} is a popular machine learning library for Python and one of the reasons for the growing popularity of python as a language for data analysis. 
%\item Spark MLlib~\citep{meng2016mllib} is the machine learning component of the rapidly growing big data framework Apache Spark that is developed with Scala.
%\item Caret~\citep{Kuhn2018} provides common interfaces for many learning algorithms implemented in R. Through this, Caret provides a harmonized Application Programing Interface (API) that allows using many learning algorithms that are implemented with different APIs as part of different R packages. Thus, Caret greatly reduces the difficulty of trying out different algorithms to solve a problem. 
%\end{itemize}

\begin{table}[]
\centering
\begin{tabular}{lllp{6cm}}
\textbf{Framework} & \textbf{Version} & \textbf{Language} & \textbf{Description} \\
\hline\hline
Scikit-learn & 1.0.1 & Python &
Scikit-learn~\citep{Pedregosa2011} is a popular machine learning library for Python and one of the reasons for the growing popularity of python as a language for data analysis.
\\
Weka & 3.8.5 & Java &
Weka is a popular machine learning library for Java which had the first official release in 1999. The library is very mature and has been used by researchers and practitioners for decades and is, e.g., part of the Pentaho business intelligence software.\tablefootnote{https://www.hitachivantara.com/en-us/products/data-management-analytics/pentaho-platform.html}
\\
Spark MLLib & 3.0.0 & Scala &
Spark MLLib~\citep{meng2016mllib} is the machine learning component of the rapidly growing big data framework Apache Spark~\cite{Zaharia2010} that is developed with Scala.
\\
Caret & 6.0-90 & R &
Caret~\citep{Kuhn2018} provides common interfaces for many learning algorithms implemented in R. Through this, Caret provides a harmonized Application Programing Interface (API) that allows using many learning algorithms that are implemented with different APIs as part of different R packages. Thus, Caret greatly reduces the difficulty of trying out different algorithms to solve a problem. 
\\
\end{tabular}
\caption{\rev{Overview of the machine learning libraries selected for our study.}}
\label{tbl:frameworks}
\end{table}

\subsection{Methodology}

\begin{figure}
\centering

\tikzstyle{block} = [draw, text width=7.5em, text centered]
\tikzstyle{line} = [draw, >=latex']

\begin{tikzpicture}
\node (SK) at (-3,3) {\includegraphics[width=2cm]{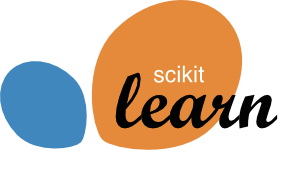}};
\node (WK) at (-1,3) {\includegraphics[width=1.5cm]{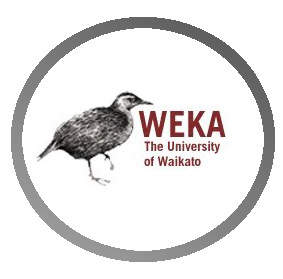}};
\node (SP) at ( 1,3) {\includegraphics[width=2cm]{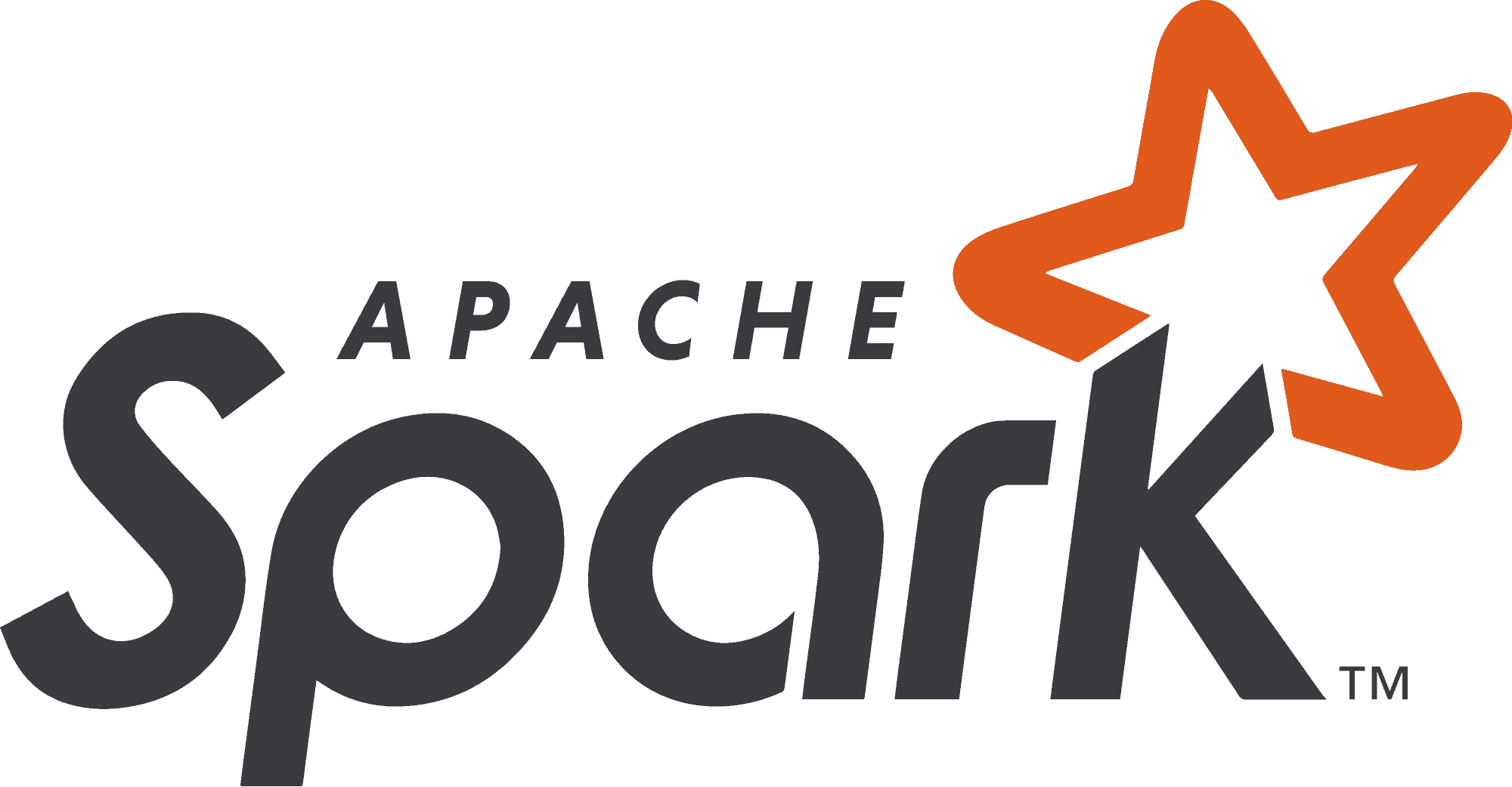}};
\node (CR) at ( 3,3) {\includegraphics[width=2cm]{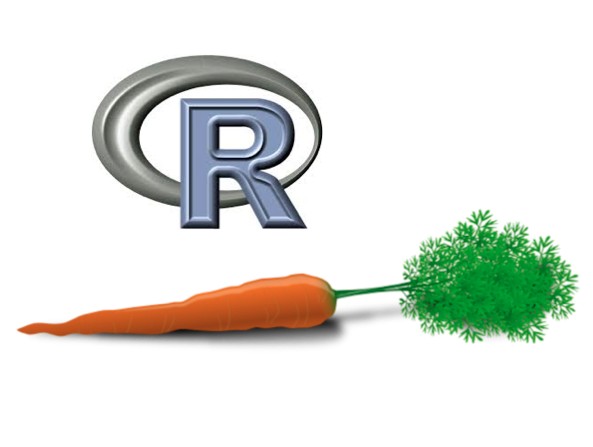}};
\node[shape=rectangle,draw=black,minimum width={9cm},minimum height={2cm}] (B1) at (0,3) {};
\node[shape=rectangle,draw=black,minimum width={9cm}] (ALG) at (0,1) {Lists of Algorithms};
\node[shape=rectangle,draw=black,minimum width={9cm}] (MATCH) at (0,-0.5) {Algorithms that are in $>3$ frameworks};
\node[shape=rectangle,draw=black,minimum width={9cm}] (CAND) at (0,-2) {Algorithms suitable for differential testing};
\node[shape=rectangle,draw=black,minimum width={9cm}] (RES) at (0,-3.5) {Results of differential testing};

\draw [-latex] (B1) -- node [midway, right ] {Collect Classification Algorithms} (ALG);
\draw [-latex] (ALG) -- node [midway, right ] {Match algorithms} (MATCH);
\draw [-latex] (MATCH) -- node [midway, right ] {Compare API documentation} (CAND);
\draw [-latex] (CAND) -- node [midway, right ] {Run differential tests} (RES);

\node[rotate=90,text depth={9.25cm},draw=black,minimum width={5cm},minimum height={10cm}] (P1) at (-0.25, 1.6) {\textit{\textbf{Phase 1}}};
\node[rotate=90,text depth={9.25cm},draw=black,minimum width={1.5cm},minimum height={10cm}] (P1) at (-0.25, -1.7) {\textit{\textbf{Phase 2}}};
\node[rotate=90,text depth={9.25cm},draw=black,minimum width={1.5cm},minimum height={10cm}] (P1) at (-0.25, -3.25) {\textit{\textbf{Phase 3}}};
\end{tikzpicture}
\caption{Methodology Overview}
\label{fig:method}
\end{figure}
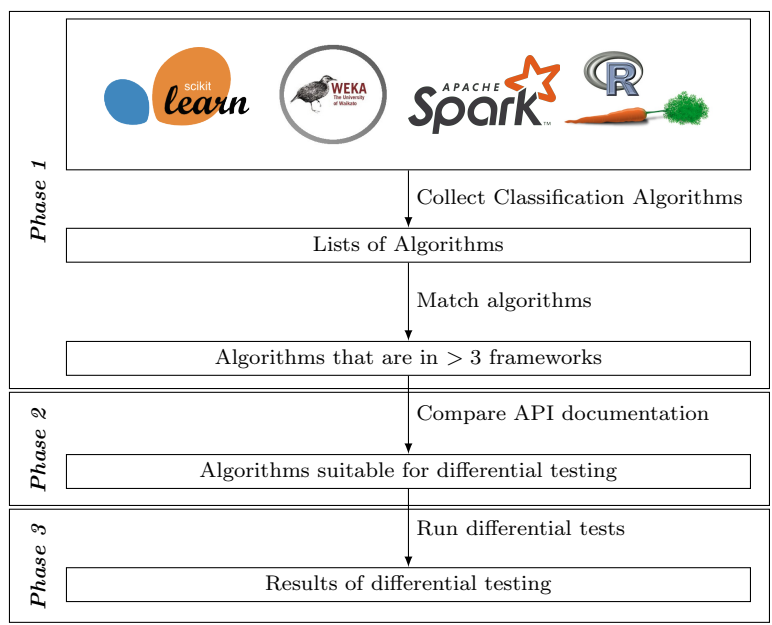

Our case study was conducted in three phases, as depicted in Figure~\ref{fig:method}. Each phase provides insights into a different aspect of the usefulness of differential testing. Moreover, each phase builds on the results of the previous phase. 

In the first phase, we evaluate the basic assumption of differential testing: we have multiple implementations of the same algorithm. To evaluate this, we analyze each framework and extract the classification algorithms. We then compare these lists to each other to determine overlaps. The number of overlaps determines the potential for the differential testing, because this means that multiple implementations are available. For all frameworks, we exclude pure meta learners from our analysis. Meta learners are wrappers around other classifiers (e.g., generic boosting~\citep{Freund1997} or bagging~\citep{Breiman1996}), i.e., they propagate the learning to other classifiers called base classifiers. Thus, meta learners require special consideration: we must first establish that there are other equal algorithms, because otherwise it would be unclear if differences observed when comparing meta learners are due to deviations between the meta learners, or due to differences between the base classifiers. 

In the second phase, we evaluate if we actually manage to implement differential tests. This is non-trivial due to the hyperparameters of algorithms. Hyperparameters allow the users of a machine learning algorithm to configure the training and/or the resulting model, e.g., by restricting sizes, enforcing regularization, or even by selecting which variant of an algorithm is used. While the same is true for deep learning, there is a big difference: with deep learning, every user of a library has to configure the hyperparameters for each neural network and training. \rev{On} the one hand, this increases the burden on the developers, as this is not pre-configured. On the other hand, there are no restrictions and thus, defining an equivalent set of hyperparameters is not problematic. This is different with other algorithms. For example, there are many different decision tree variants, such as CART~\citep{Brieman1984}, ID3~\citep{quinlan1986induction}, and C4.5~\citep{Quinlan1993}. A random forest could be based on each of these variants, in which case the random forests could not be directly compared. However, even the same variant of a decision tree is not always the same, as concrete implementations may not implement the original specification or they may have added more features, e.g., for avoiding overfitting through limiting tree depths or requiring a certain number of instances to make decisions. Thus, within the second phase we compare different implementations based on their API documentations and determine for which overlaps we manage to find matching configurations. For Caret, we also included the API documentation of the R packages that are providing the implementation that is wrapped by Caret in our analysis. Furthermore, our analysis also includes the reasons for cases where we fail to match the hyperparameters. 

In the third phase, we implement a simple differential testing approach. We use four data sets as input. 

\begin{itemize}
    \item UNIFORM: \rev{200 randomly generated instances}, with ten features that are uniform in the interval $[0,1]$ with two classes, such that 50\% of the data are in each class. The separation between the classes is based on a 10-dimensional rectangle, i.e., a clear separation between the classes is possible. 
    \item RANDOM: \rev{200 randomly generated instances}, where the features are uniform within the interval $[0,1]$ with two classes that are randomly assigned. Thus, no reasonable classification is possible. 
    \item BC: The \emph{breast cancer} data from the UCI repository \rev{has 569 instances 30 numeric features for cancer prediction}. The features are normalized to the interval $[0,1]$.
    \item WINE: The \emph{wine} data from the UCI repository \rev{has 178 instances with 11 numeric features for wine quality prediction}. The features are normalized to the interval $[0,1]$.
\end{itemize}

The first two data sets are reused from our prior work \citep{Herbold2022} for the smoke testing of algorithms. With UNIFORM, we can test if different implementations perform the same, for data that is relatively easy to learn based on informative features. With RANDOM, we test if the implementations also yield the same results, in case the data is not informative. This scenario can be considered ``more difficult'', as the resulting model is more or less random. Still, the ``randomness'' should depend on the algorithm and, ideally, different implementations still yield the same results. The other two data sets are popular data sets that are often used as examples for using machine learning models. All data sets are randomly split into 50\% training data and 50\% test data. 

The data we use have two aspects in common: they have two classes and the features are within well-bounded and numerically not problematic feature spaces. Consequently, we are implementing a relatively simple differential testing scenario, i.e., we do not intentionally stress the algorithms.\footnote{This is also the reason why we only use the test cases UNIFORM and RANDOM from our prior work~\citep{Herbold2022}. \rev{Our analysis of crashes showed that such data does not lead to crashes, which would make the analysis of the effectiveness of the differential testing more difficult.}} Our rationale for this is that we want to test the general feasibility of differential testing to find errors due to deviations between implementations. If our data would target corner cases, our results could overestimate the effectiveness \rev{because differences might be larger for extreme data} or be polluted by algorithms crashing failures due to numeric issues. \rev{Moreover, classification with more classes either uses exactly the same algorithms (e.g., decision trees and neural networks) or a combination of multiple binary models (e.g, one-vs-follower approach with support vector machines). However, we may also underestimate the difference between algorithms for multiple classes.}

We use four criteria to evaluate the outcome of the differential tests for two algorithms $A^1$ and $A^2$. The first two criteria measure if the results are equal, i.e., if two different implementations yield the same results. To evaluate this, we consider the number of differences between the classes \rev{predicted by the two algorithms} defined as 
\begin{equation}
\Delta = \sum_{i=1}^n \mathds{1}_{f^1(x_i)\neq f^2(x_i)}
\end{equation}
as the first criterion and the number of deviations between the scores defined as
\begin{equation}
\Delta_{score} = \sum_{i=1}^n \mathds{1}_{f_{score}^1(x_i)\neq f_{score}^2(x_i)}
\end{equation}
as the second criterion. Please note that we consider scores to be equal, if the difference is less than 0.001. Smaller differences are almost always irrelevant and may easily be due to differences with how floating point operations are implemented. If we were to enforce exact equality of scores, this could inflate the false positive test results. 

Additionally, we often cannot expect that two results of different implementations are exactly the same, e.g., because an aspect of the algorithm is randomized. However, the results should not differ too much and we should still get roughly the same result in terms of overall performance. We evaluate this by looking at the significance of the differences between the classes and scores and derive the two remaining criteria. The third criterion considers the differences in the scores. Since the scores are continuous values and we cannot make general assumptions regarding their distribution, we use the Kolmogorov-Smirnoff test. Thus, we check if the scores of the two models have the same distribution. The fourth criterion applies this concept to the classes. Similarly, we use the Chi-Squared test to compare the classes to check if the differences in classifications are significant. We reject the null hypotheses of the tests (no difference) if we observe a p-value of less than 0.05. We do not conduct any correction of repeated tests, as this correction would be dependent on the number of tests that are executed, i.e., become a parameter of the test campaign. This is hard to implement in practice. However, this means that we expect that about 5\% of the significant differences we observe to be false positives, because the p-value follows a uniform distribution if the null hypothesis is true. If and how this affects the differential testing will be discussed based on our results. 

\begin{figure}
\centering
\includegraphics[width=\textwidth]{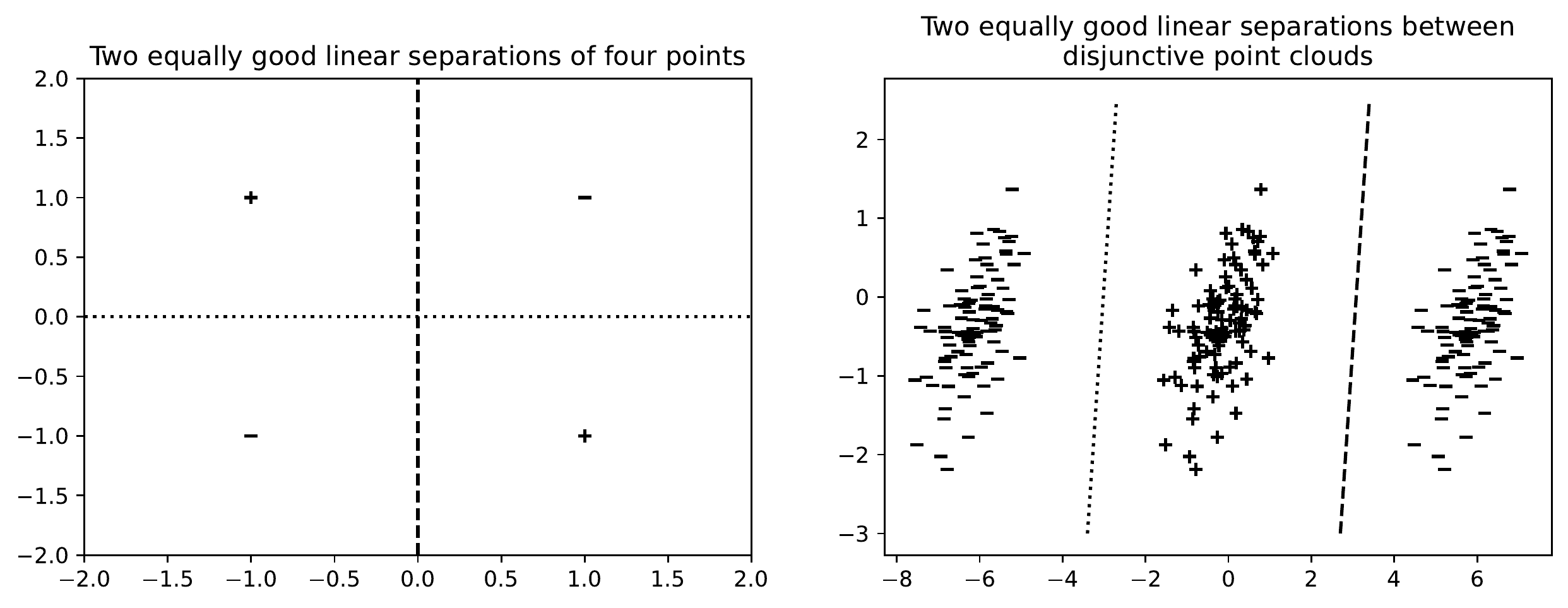}
\caption{Two examples of data sets, where the dotted and dashed lines represent possible linear separations of the data \rev{that could, e.g., be computed by a logistic regression}. Both models are equally good \rev{in terms of accuracy.}}
\label{fig:differences-example}
\end{figure}

While it may seem counterintuitive that we use four criteria to evaluate the effectiveness of software tests, these criteria allow us not only to answer the general question if differential tests are effective, but also how lenient the comparison of the results must be in order to not yield false positives. This is due to the nature of machine learning. Figure~\ref{fig:differences-example} shows two cases, where correct implementations of the same algorithm can yield different \rev{optimal} outcomes. In both examples, the absolute differences we observe (the first two criteria) are not suitable to determine that a test failed, i.e., one result is wrong. However, the differences between the outcomes should not be statistically significant. Ideally, we hope to see the limits of both absolute comparisons (criteria one and two) and statistical comparison (criteria three and four) for the differential testing as criteria for passing or failing differential tests. 

Additionally, we apply all four criteria to both the training and the test data. Both training and test data have advantages and disadvantages. The training data has the advantage that it is readily available. However, software tests that only rely on training data may miss cases in which the implementations does not correctly generalize beyond the training data. This is the advantage of test data, as the computed functions for scoring and classification are evaluated on an independent data set. However, this may also be prone to more false positives, because equally good results on the training data may lead to different results on the test data (see Figure~\ref{fig:differences-example}). \rev{Since we use the same amount of training and test data, the likelihood of differences between implementations on the training and test data is not impacted by the amount of data available.}

\SetKwComment{Comment}{// }{\;}
\DontPrintSemicolon
\RestyleAlgo{ruled}
\begin{algorithm}
\caption{Differential testing algorithm}\label{alg:two}
\KwIn{$A^1$, $A^2$, $(X^{train}, Y^{train})$, $(X^{test}, Y^{test})$}
\KwOut{Number of differences $\Delta^{train}, \Delta^{test}, \Delta_{score}^{train}, \Delta_{score}^{test}$\ and $p$-values $p_{KS}^{train} , p_{KS}^{test}, p_{\chi^2}^{train}, p_{\chi^2}^{test}$}
\Comment{Run training}
$f^1, f^1_{score} \gets A^1(X^{train}, Y^{train})$\;
$f^2, f^1_{score} \gets A^2(X^{train}, Y^{train})$\;
\BlankLine
\Comment{Criterion 1: Absolute differences between classes}
$\Delta^{train} \gets \sum_{i=1}^{n^{train}} \mathds{1}_{f^1(x^{train}_i)\neq f^2(x^{train}_i)}$\;
$\Delta^{test} \gets \sum_{i=1}^{n^{test}} \mathds{1}_{f^1(x^{test}_i)\neq f^2(x^{test}_i)}$\;
\BlankLine
\Comment{Criterion 2: Absolute differences between scores}
$\Delta_{score}^{train} \gets \sum_{i=1}^{n^{train}} \mathds{1}_{f^1_{score}(x^{train}_i)\neq f^2_{score}(x^{train}_i)}$\;
$\Delta_{score}^{test} \gets \sum_{i=1}^{n^{test}} \mathds{1}_{f^1_{score}(x^{test}_i)\neq f^2_{score}(x^{test}_i)}$\;
\BlankLine
\Comment{Criterion 3: Significance of difference between scores}
$p_{KS}^{train} \gets KS.test(f^1_{score}(X^{train}), f^2_{score}(X^{train}))$\;
$p_{KS}^{test} \gets KS.test(f^1_{score}(X^{test}), f^2_{score}(X^{test}))$\;
\BlankLine
\Comment{Criterion 4: Significance of difference between classifications}
$p_{\chi^2}^{train} \gets \chi^2.test(f^1(X^{train}), f^2(X^{train}))$\;
$p_{\chi^2}^{test} \gets \chi^2.test(f^1(X^{test}), f^2(X^{test}))$
\end{algorithm}

\subsection{Phase 1: Overlap of Algorithms}

Table~\ref{tbl:overlap} shows the overlapping algorithms of the four machine learning frameworks we found using the API docs. We grouped the algorithms by their underlying paradigm, e.g., \textit{Naive Bayes} or \textit{Decision Tree}. The first row demonstrates the reason for this. While Scikit-learn provides unique classes for different variants of Naive Bayes, Spark MLlib has only a single class. However, the implementation in Spark MLlib can be configured to use different variants, e.g., Gaussian Naive Bayes or Multinomial Naive Bayes. Our data shows that there is a significant overlap: 
\begin{itemize}
    \item Naive Bayes, Decision Trees, Random Forest, Support Vector Machines, Multilayer Perceptrons, and Logistic Regression can be found in all frameworks;
    \item a trivial classifier, $k$-Nearest Neighbor, and Gradiant Boosting in at least three of the frameworks; and
    \item a Perceptron, Stochastic Gradient Descent, Gaussian Process, LDA, QDA, Nearest Centroid, Extra Trees, and Logistic Model Trees are present in two frameworks.
    \item Additionally, there are 47 algorithms which are only implemented by a single framework (see Table~\ref{tbl:overlap-full} in the appendix).
\end{itemize}

While we are not aware of any scientific data regarding this, the eight algorithms that can be found in all frameworks are all well-known and frequently mentioned in lists of techniques that should be learned.\footnote{Some examples from Kaggle.com: \url{https://www.kaggle.com/general/253858} \url{https://www.kaggle.com/getting-started/96750} \url{https://www.kaggle.com/getting-started/268971}} There seems to be a link between the potential for differential testing and the popularity of approaches. Intuitively, this makes sense, because developers of machine learning frameworks likely implement popular methods first. For example, Scikit-learn has a popularity/maturity based inclusion criterion based on the number of citations.\footnote{\url{https://scikit-learn.org/stable/faq.html\#what-are-the-inclusion-criteria-for-new-algorithms}} Another interesting aspect is that there are sometimes multiple implementations for the same algorithm within a single framework. For Scikit-learn, Weka, and Spark MLlib, these are different variants of an algorithm.\footnote{For every rule, there is an exception, in this case the RidgeClassifier from Scikit-learn, which is a special case of the LogisticRegression.} For Caret, this is not the case. The reason for this is that Caret is a meta framework, i.e., a framework that wraps the implementations of different R packages to provide a common API. Due to this, there are sometimes implementations from different R packages for the same algorithm, e.g., in the first row from the packages \texttt{naive\_bayes} and \texttt{nb}. Consequently, there is even a possibility for differential testing within the Caret package, without even using other frameworks. 

\begin{mdframed}
There is a large potential for differential testing beyond deep learning. This potential is strongest for well-known and popular algorithms, while newer or specialized algorithms are not good candidates for differential testing. 
\end{mdframed}

\begin{table}[]
\centering
\begin{adjustbox}{angle=90}
\begin{tabular}{p{4.5cm}p{3.3cm}p{3cm}p{6.9cm}}
\textbf{Scikit-learn} & \textbf{Weka} & \textbf{Spark MLlib} & \textbf{Caret}  \\
\hline\hline
BernoulliNB, ComplementNB, MultinomialNB, GaussianNB, CategoricalNB & NaiveBayesMultinomial, NaiveBayes & NaiveBayes & awnb, naive\_bayes, nb, manb
\\\hline
DecisionTreeClassifier & SimpleCart, J48, DecisionStump, HoeffdingTree & DecisionTreeClassifier & C5.0Tree, C5.0Rules, rpartCost, C5.0Cost, rpartScore, rpart, rpart1SE, rpart2, C5.0, ctree, ctree2, bartMachine
\\\hline
RandomForestClassifier & RandomForest & RandomForestClassifier & wsrf, RRF, RRFglobal, rfRules, ordinalRF, ranger, Rborist, rf, ORFlog, ORFpls, ORFridge, ORFsvm, cforest
\\\hline
LinearSVC, NuSVC, SVC & SMO & LinearSVC & svmSpectrumString, svmRadial, svmRadialCost, svmRadialSigma, svmPoly, svmLinear, svmLinear2, svmExpoString, svmRadialWeights, svmBoundrangeString, svmLinearWeights, lssvmRadial, lssvmPoly, lssvmLinear, svmLinear3, svmLinearWeights2
\\\hline
MLPClassifier & MultilayerPerceptron & MultilayerPerceptron-Classifier & mlpKerasDecayCost, mlpKerasDecay, mlpKerasDropoutCost, mlpKerasDropout, mlpSGD, mlpML, mlpWeightDecayML, mlp, mlpWeightDecay, monmlp
\\\hline
LogisticRegression, RidgeClassifier & Logistic, SimpleLogistic & LogisticRegression & regLogistic, plr, polr
\\\hline
DummyClassifier & ZeroR &  & null
\\\hline
KNeighborsClassifier & IBk &  & snn, kknn, knn
\\\hline
GradientBoostingClassifier &  & GBTClassifier & gbm\_h2o, gbm
\\\hline
Perceptron & VotedPerceptron &  & 
\\\hline
SGDClassifier & SGD &  & 
\\\hline
GaussianProcessClassifier &  &  & gaussprRadial, gaussprPoly, gaussprLinear
\\\hline
LinearDiscriminantAnalysis &  &  & slda, sparseLDA, sda, rrlda, Linda, rlda, rda, PenalizedLDA, pda, pda2, Mlda, loclda, stepLDA, lda, lda2, RFlda, dda
\\\hline
NearestCentroid &  &  & pam
\\\hline
QuadraticDistriminantAnalysis &  &  & qda, QdaCov, stepQDA
\\\hline
ExtraTreeClassifier &  &  & extraTrees
\\\hline
 & LMT &  & LMT
\\\hline
\end{tabular}
\end{adjustbox}
\caption{Overlapping algorithms between frameworks based on a scan of algorithm names in the API docs. For Scikit-learn, Weka, and Spark MLlib we report the names of the classes, in which the algorithms are implemented. For Caret, we report the name of the method within Caret.}
\label{tbl:overlap}
\end{table}

\subsection{Phase 2: Feasible Subset}

Table~\ref{tbl:feasible} reports the feasible combinations of algorithms that we believe should be the same, based on the API documentation. Overall, we identified three variants of Naive Bayes (GNB, KDENB, MNB), two variants of the random forest (RF1 with fixed depth and RF2 without fixed depth), three variants of Support Vector Machines (LSVM with a linear kernel, PSVM with a polynomial kernel, and RBFSVM with a RBF kernel), the Multilayer Perceptron (MLP), the trivial classifier (DUMMY), the $k$-Nearest Neighbor (KNN) algorithm, and three variants of Logistic Regression (LR without regularization, RIDGE with ridge regularization and LASSO with lasso regularization). 

This means we found feasible combinations for all candidates with implementations in at least three of the four frameworks, with the exception of the gradient boosting trees. The reason we failed here is because the Spark MLlib API does not specify which kind of decision tree is implemented. Even beyond this, while we identified many different decision trees when analyzing the potential, we could not find any feasible combination of implementations. For CART this almost worked, but was not possible because Caret did not allow passing the appropriate parameters to the underlying library. Another issue was the options that were configurable through hyperparameters (incl. defaults) and hard coded values. In all cases, we needed to carefully consider the hyperparameters to set them to appropriate values to behave the same as with the other frameworks. There was no case, where the out-of-the box default parameters between two implementations were equal for all implementations. As part of this analysis, we tried to identify hyperparameters, that can be set with all implementations and that should have the same effect. We tried to maximize the number of such equal hyperparameters, but often only found few, if any. In case the hyperparameters were not available in all frameworks, we tried to determine which value was used by the implementation from the API. Often, this was impossible because such implicit assumptions on the algorithm configuration that are not exposed to users through the API as configurable hyperparameters are also not documented. Overall, this means that the feasible subsets do not cover the full capabilities of the frameworks, i.e, it is not possible for any of the algorithms to test the full set of hyperparameters within our case study. 

A notable aspect of finding feasible combinations is that a good knowledge about the algorithms is required. For example, the different implementations of the RIDGE group use different variants to define the strength of the regularization. While the Scikit-learn implementation LogisticRegression uses the inverse regularization strength $\frac{1}{2\cdot\alpha}$, all other implementations, including the Scikit-learn implementation RidgeClassifier directly use $\alpha$ as regularization strength. Without understanding the underlying mathematical principles, deriving suitable tests is not possible. However, we do not believe that this is, in practice, a restriction on the feasibility of such tests: developers who actively work on such algorithms, either through quality assurance or development, should have the required background on the techniques anyways. 

\begin{mdframed}
The potential for differential testing can often be operationalized in form of feasible combinations that should yield the same behaviour. However, such tests can only cover a subset of the implemented functionality, because of differences between the configuration options that are exposed as hyperparameters. 
\end{mdframed}

\begin{table}
\renewcommand{\arraystretch}{1.5}
\begin{adjustbox}{angle=90}
\begin{tabular}{lp{3cm}lll}
\textbf{Group} & \textbf{Scikit-learn} & \textbf{Weka} & \textbf{Spark MLlib} & \textbf{Caret} \\
\hline
GNB & GaussianNB & NaiveBayes & NaiveBayes & naive\_bayes, nb \\
KDENB & & NaiveBayes & & naive\_bayes, nb \\
MNB & MultinomialNB & NaiveBayesMultinomial & NaiveBayes & \\
RF1 & RandomForestClassifier & RandomForest & & ranger, rborist \\
RF2 & RandomForestClassifier & RandomForest & & ranger, rborist, rf \\
LSVM & SVC & SMO & LinearSVC & svmLinear, svmLinear2, svmLinear3 \\
PSVM & SVC & SMO & & svmPoly \\
RBFSVM & SVC & SMO & & svmRadial \\
MLP & MLPClassifier & MultilayerPerceptron & MultilayerPerceptronClassifier & mlp*, mlpSGD* \\
DUMMY & DummyClassifier & ZeroR & & null \\
KNN & KNeighborsClassifier & IBk & & knn \\
LR & LogisticRegression & Logistic & LogisticRegression & regLogistic, plr, polr \\
RIDGE & LogisticRegression, RidgeClassifier & Logistic & LogisticRegression & regLogistic, plr \\
LASSO & LogisticRegression & & LogisticRegression & regLogistic \\
\end{tabular}
\end{adjustbox}
\caption{Feasible combinations with at least three implementations. Parameters are ommited and can be found in Table~\ref{tbl:feasible-full} in the Appendix. The mlp classifiers of Caret could not be executed because they require an R package which is not on CRAN anymore.} \label{tbl:feasible}
\end{table}

\subsection{Phase 3: Test Execution}

We executed the tests for all pairs of algorithms we identified in the second phase, i.e., 87 pairs of algorithms with the parameters defined in Table~\ref{tbl:feasible-full} in the appendix on the four data sets with both training and test data for evaluation, i.e., we have $87~\text{pairs} \cdot 4~\text{data sets} \cdot 2~\text{training/test data} = 696$ comparisons. We observed difference between the classifications for 457 pairs (65.6\%), 67 (9.6\%) differences were significant. Since some algorithms cannot compute scores,\footnote{All algorithms in the groups LSVM, PSVM, RBFSVM, the algorithm ranger from the RF1 and RF2 groups, and the algorithm RidgeRegression from the RIDGE group.} we only have 432 pairs with scores. We observed differences between the scores for 313 pairs (72.4\%), 141 (32.6\%) were significant. For better insights, we looked at the results for each group of algorithms in detail. Table~\ref{tbl:test-results} summarizes the results for each group. We refer to the algorithms by their framework. In case there are multiple implementations per framework, we also provide the name of the implementations. 

\begin{table}
\renewcommand{\arraystretch}{1.5}
\centering
\begin{tabular}{l p{9.5cm}}
\textbf{Group} & \textbf{Summary of test results} \\\hline
\textbf{GNB} & Classes are equal or almost equal, with small differences for Weka. Scores are almost always different, except for the Caret-naive\_bayes and Caret-nb, and the Spark MLlib and Scikit-learn implementations, which are equal. No differences are significant. \\
\textbf{KDENB} & Small but insignificant differences in classes and scores between the Caret-naive\_bayes and Caret-nb. Large differences to Weka, that are significant for both classes and scores on about half of the data sets. \\
\textbf{MNB} & Classes and scores are equal, with one exception: the Scikit-learn implementation sometimes classifies one instance differently. These differences are in cases where the score is almost exactly 0.5 and for one framework slightly smaller and for the other framework slightly larger than 0.5, e.g., for Scikit-learn 0.499 and for Weka 0.501. \\
\textbf{RF1} & Results are never equal, with larger differences on test than on training data. The differences between classes are not significant, except in two cases on the RANDOM data, where the classes are significantly different. The differences between the scores are almost always significant. \\
\textbf{RF2} & Results are never equal, with larger differences on test than on training data. The differences of the classes are not significant, except in two cases on the RANDOM data. The scores between Weka and Caret-rboist is only significant once on the WINE data. The scores of Scikit-learn are significantly different most of the time.\\
\textbf{LSVM} & Results are never equal, with mostly small and insignificant differences. The exception is Spark MLlib, which has large difference to the other implementations that are significant on the RANDOM and WINE data.\\
\textbf{PSVM} & Scikit-learn and Weka are equal. Caret yields different results, but these differences are only significant on the UNIFORM data. \\
\textbf{RBFSVM} & Scikit-learn and Weka are equal. Caret yields different results and these differences are almost always significant. \\
\textbf{MLP} & On the RANDOM and UNIFORM data, the classes are almost equal between all implementations with no significant difference. On the WINE and BC data there are large and significant differences, where Weka disagrees with the other frameworks. The differences between scores are always large and significant. \\
\textbf{DUMMY} & The classes are always equal, the scores depend on the implementation of the trivial model: Caret and Weka have the same approach, Scikit-learn always disagrees.\\
\textbf{KNN} & The classifications are equal or almost equal between all implementations. The scores of Scikit-learn and Caret are also equal or almost equal, with the exception of the WINE data. Here, Caret is equal to Weka instead. On the other data sets, Weka has a large and significant difference from the other implementations.\\ 
\textbf{LR} & The classes are equal or almost equal. The scores are almost always equal, except on the BC data, where we observe significant differences between all implementations. \\
\textbf{RIDGE} & The classes and scores of Weka, Scikit-learn-LogisticRegression and Caret-plr are equal. The other implementations have small and insignificant differences for the classes and large and significant deviations for the scores.\\
\textbf{LASSO} & The classifications of Caret and Scikit-learn are equal or almost equal. Spark MLlib has large differences, but they are only significant on the RANDOM data. The scores between all implementations are different. The differences between Caret and Scikit-learn are mostly not significant, Spark MLlib is always significantly different. 
\end{tabular}
\caption{Overview of the results of the execution of the differential tests.}
\label{tbl:test-results}
\end{table}

\begin{figure}
\centering
\includegraphics[width=\textwidth]{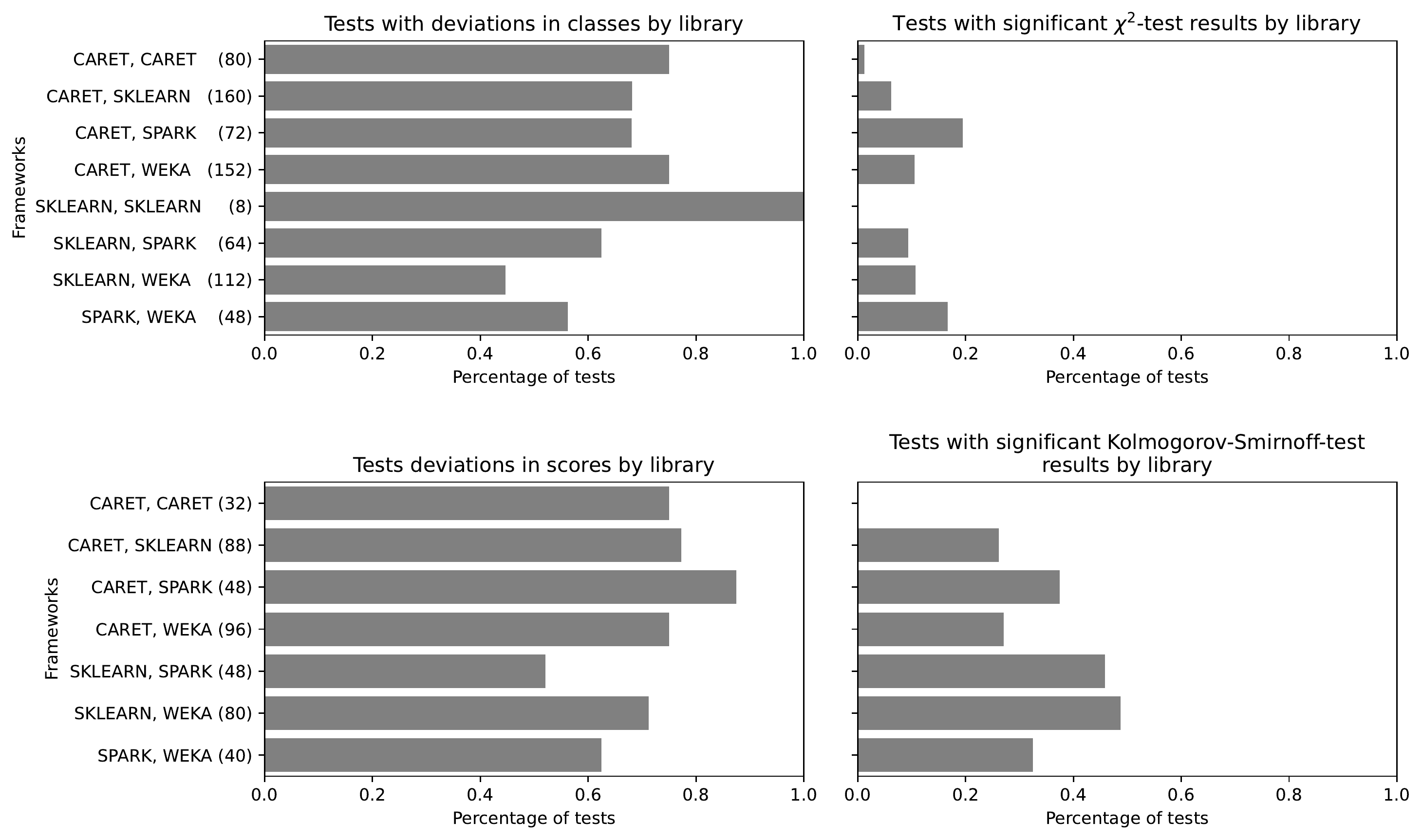}
\caption{\rev{Differences observed for pairs of libraries. The labels show the pair of libraries and the number of tests executed. For example, ``CARET, SKLEARN (160)'' means that we ran 160 tests where one classifier was from either framework.}}
\label{fig:results_by_framework}
\end{figure}

\begin{figure}
\centering
\includegraphics[width=\textwidth]{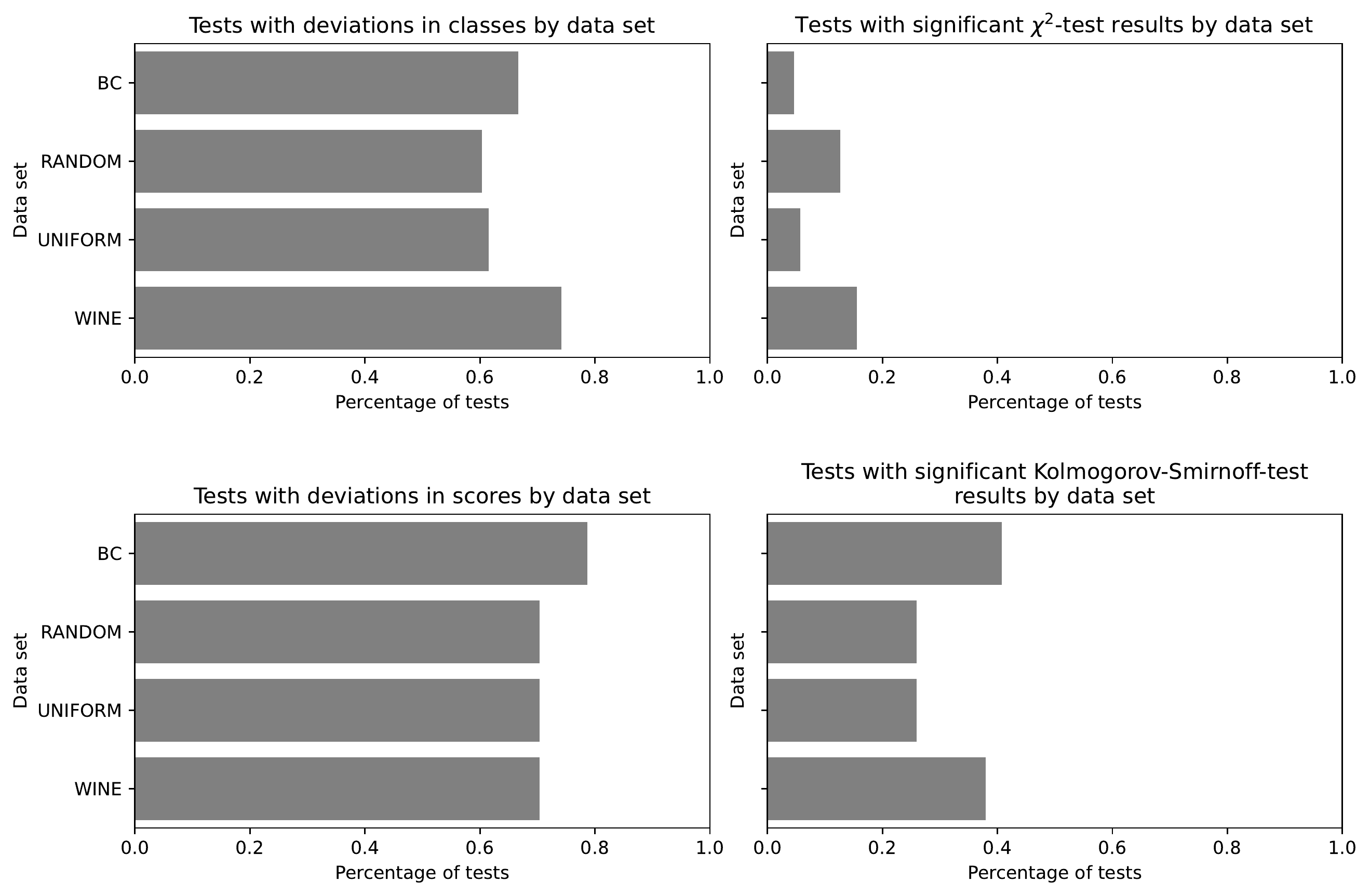}
\caption{\rev{Differences observed with the data sets we used.}}
\label{fig:results_by_dataset}
\end{figure}

Overall, these results paint a complex picture regarding the effectiveness. For example, we found no patterns regarding which frameworks often disagree with each other, \rev{except for a tendency that the deviations between different implementations wrapped by Caret are mostly not significant (Figure~\ref{fig:results_by_framework}). We also found no pattern regarding} which data sets are responsible for the deviations \rev{(Figure~\ref{fig:results_by_dataset})}. Moreover, there is no group of implementations in which both classes and scores are equal. We observe so many absolute differences that we cannot really conclude anything regarding concrete bugs. Nevertheless, we found a couple of repeating aspects. 

\begin{itemize}
\item Perhaps unsurprisingly, non-deterministic algorithms like random forests or MLPs can only be compared using statistical significance. This seems to work well for the classifications using the $\chi^2$ test. However, our results indicate that the scores should not be compared, as the differences are very often significant.
\item Perhaps surprisingly, we often observe different classifications with deterministic algorithms, even to the point where we cannot state which algorithms are likely correct. For example, the small differences for GNB makes it impossible to determine, without a detailed triage of all involved algorithms, to infer if any of the implementations contain a bug causing the small deviations, or if this is natural due to implementation choices. Most of the time these differences are not statistically significant, which further raises the question if investigating these issues would be worth the effort. 
\item Scores often depend on the implementation. Even for a ``simple'' algorithm like LR, we observe significant differences. Thus, we cannot recommend to use differential testing for scores, as there is a very large risk of false positives. 
\end{itemize}

We also note that our choice to use a threshold for the p-value without correction is not without consequences. For the Kolmogorov-Smirnoff tests between the scores, we have over 32\% results in which we observe significant, i.e., far more than the expected 5\% of false positives. Since there are too \rev{many} differences between scores anyways, the impact of possible false positives is limited. For the Chi-Squared, we have only about 10\% significant difference. Consequently, the randomness of the p-value distribution if the null hypothesis is true would explain about half of the significant differences we observe. This adds another wrinkle to our already hard to interpret results. However, we would argue that for deterministic algorithms, only very small differences are acceptable, and anything large enough to be picked up as a significant signal by the Chi-Squared test actually points towards a significant difference between deterministic implementations. For randomized algorithms, a potential solution would be to use repeated tests, e.g., 100 different UNIFORM samples. One could then observe if the number of significant deviations is less than the expected false positive rate of 5\%, which would indicate acceptable differences, or if it is large which would indicate a significant difference. 

Due to these issues, we believe that differential testing may be used by developers who already know the details of their implementation, to compare their work with other implementations. \rev{However, whether this is more effective and/or efficient than directly only debugging a single implementation is not clear without further study.} Using differential testing as pseudo-oracle~\citep{Davis1981} to automatically drive test campaigns between tools seems to have only a limited potential. 

\begin{mdframed}
The results of the tests indicate that an effective and efficient usage of differential testing between implementations is mostly not possible on an absolute level, i.e., expecting equal results. Tests for significant differences of classifications seem to be the only feasible option, but should only be used with proper care regarding false positive and require that tester are already experts for the implementations under test. 
\end{mdframed}

\section{Discussion}
\label{sec:discussion}

Overall, our results show that while we have a large potential, it is very difficult to find feasible configurations of the algorithms. When we then execute these tests, we observe that there are many differences, to the point where we cannot distinguish between possible problems and normal deviations. We saw that especially the scores depend a lot on the implementation. Even if we just consider whether the binary classifications are equal, we observed too many differences. \rev{There is no reason to believe that this problem should not exist for more classes as well.} While we expected that there would be some differences between the frameworks, the magnitude of these differences surprised us. We note that the data sets we use are neither very large, nor known to be numerically challenging. 

Consequently, we believe that differential testing is just not possible between frameworks, unless a very lenient test oracle is used that avoids the large amount of false positives, e.g., using the Chi-Squared tests. However, this comes at the cost of a larger chance of false negatives as well, because unexpected small differences between classes and any kind of difference between scores cannot be detected. The cases where this would also be possible are rather the exception then the rule. We can only speculate regarding the reasons for this. However, we believe that there are four main drivers for the differences we observe: 

\begin{itemize}
    \item The definition of the algorithms in the literature may not be sufficiently detailed to guarantee that everything is implemented in exactly the same way. While we believe that this could sometimes be a reason, this cannot explain all differences we observe. For example, LSVM have a clear mathematical definition which is well known and should lead to the same optimal result, except in few corner cases. In comparison, no two implementations yield exactly the same results. 
    \item It is possible that the API documentations are incomplete and there are hidden values of hyperparameter that are hard-coded in the implementations but not visible through the documentation. However, in this case we should see clearer patterns in the deviations, e.g., one algorithm deviating from the others on all data sets. While we sometimes observe this, there are relatively few cases were only one implementation disagrees with all others. Usually, many implementations disagree with each other within one group.
    \item Optimizations for the numerical stability, run-time, or memory efficiency could lead to subtle changes in the algorithms which could explain the deviations between the implementations. The pattern of deviations should be the same as before, i.e., one algorithm disagrees with the others, which we often do not observe very often. 
    \item The different programming languages, including the libraries available, could also have an impact. While, e.g., the different pairs for GNB indicate that this may be due to such an effect, we also observe many cases that indicate that this is not an issue. For example, the different R implementations that Caret offers often disagree with each other as well, even though they are, presumably, based on the same native R features. Hence, while this could be a factor, we believe that this only plays a minor role. 
\end{itemize}

Unfortunately, we could only determine how often the above factors are the reason for differences through an extremely time-consuming manual analysis of the source code of the machine learning libraries, possibly even including libraries on which they are based, e.g., to understand the numerics involved. Such a large scale comparative audit is not possible with our resources, because we would basically need to achieve similar skills as the core developers for each of the four frameworks. 

Regardless of our lack of evidence, we believe that all of the above reasons provide a reasonable explanation for the results of our experiments. We believe that each of the reasons is a driver for some of the cases in which we observe differences, especially those where we do not observe significant differences. Unfortunately, this also means that the differential testing is not effective as automated oracle, because we would need to invest a huge amount of effort to conduct in-depth analysis of differences between most pairs of algorithms, which is not feasible. \rev{Consequently, this shows the limited capability of differential testing to provide an exact ``specification'' through a second implementation that provides a reliable oracle for deciding the correctness of software tests.}

We also believe that our results are not contradicting that differential testing works well with deep learning frameworks, e.g., PyTorch~\citep{NEURIPS2019_9015} and TensorFlow~\citep{tensorflow2015-whitepaper}: exact equality is not expected, due to the randomized nature of the training, equality of scores is also not expected because the decision boundaries of neural networks can also easily slightly change with different initialization. Thus, there is a built-in uncertainty of the training of neural networks which means that the differential testing is only concerned with checking if the classification stays roughly the same, which we also found to work. While these differences might be made slightly worse by the implementation differences or similar as we describe above, they do not hinder the differential testing, because exact equality is not the expectation, anyways. 

This is also in line with the fairly randomized algorithms we have within our data set, i.e., the random forest and the MLP. Thus, one can argue that the differential testing works beyond deep learning, if the algorithms are randomized and the only expectation is that the differences between classifications are not significant. However, subtle bugs that only affect a small amount of data can likely not be detected this way. For the algorithms in our data that are deterministic and often implement a formula or optimization approach, we should be able to observe a stronger type of equality, i.e., exactly the same results or even scores, which is not the case \rev{due to the reasons listed above, which in the end mean that the specification of the algorithms are imprecise. It follows that differential testing requires either a relaxed view on the expected equality or stronger specifications of the behavior of machine learning algorithms}. Consequently, we answer our research question as follows: 

\begin{mdframed}
While there is a large potential and it is possible to derive feasible tests, exhaustive testing of all parameter combinations or algorithms is not possible. More importantly, due to many false positive differences, the differential testing does not lead to a suitable automated test oracle if exact or almost exact equality is expected. For randomized algorithms, where some differences are expected anyways, differential testing can still be effective, but only if bugs cause large differences. 
\end{mdframed}

\section{Threats to validity}
\label{sec:threats}

We report the threats to the validity of our work following the classification by \cite{Cook1979} suggested for software engineering by \cite{Wohlin2012}. Additionally, we discuss the reliability as suggested by \cite{Runeson2009}. 

\subsection{Construct Validity}

There are crucial design choices in each phase of of our case study that could affect our results. We describe these threats below, including why the believe that the alternatives would not lead to a better construct. 

The construct of the first phase of our study is based on a coarse grouping of algorithms by the general approach, e.g., grouping decision trees together. While we have a good reason for this, i.e., that some algorithms have different variants in one implementation, while other have these variants over multiple implementation, this construct may lead to an overestimation of the potential. An indicator that this happens to some degree are the decision trees, where we have a huge group for the potential, but no feasible variants. The alternative design choice would have been a more fine-grained grouping (i.e., by concrete algorithm variant) and then possibly repeating the same implementation multiple times. However, we believe that this option has one major drawback, i.e., it would lead to merging aspects of the second phase of our study with the first phase. The more fine-grained grouping is only possibly by studying the hyperparameters through the API documentation, which is precisely what we do in the second phase, where we not only achieve a more fine-grained grouping that accounts for algorithm variants, but at the same time check if the variants are sufficiently similar to provide equal sets of hyper parameters. 

The construct of the second phase of our study is based on studying the API documentation of algorithms to find equal configurations. While we believe that this is the only practically feasible choice, there is a certain risk involved with relying on the documentation. If the documentation does not document all parameters or fails to describe all relevant aspects, we may miss feasible candidates. The alternative would be to directly check the source code to understand all options (incl. hard-coded choices). However, this would essentially be the same as an in-depth code review of the implementations under test, which we believe is not realistic, neither for us as researchers as part of a study, nor for practitioners as part of a test campaign. 

The construct of the third phase of our study is based on four different comparisons that consider equality of scores and classes, as well as significance of differences. While we believe that these are the reasonable criteria to determine if a differential test fails, other options might also work, e.g., considering the mean absolute deviation of scores. However, such an approach has the drawback that at threshold is required that defines how much deviation is acceptable. We believe that our approach based on statistical tests achieves a similar result, but with a stronger mathematical foundation. \rev{Another potential issue with our construct is that implementations may also randomly lead to unstable results, even with the same data and hyperparameters (including a fixed random seed). To account for this, we asserted that the results remain the same with ten randomly generated data sets for all algorithms. We did not observe any deviations between the classes and the scores.}

\subsection{Internal Validity}

We do not believe that there are threats to our conclusions regarding the potential, feasibility, or effectiveness, because these conclusions are directly derived from our quantitative evidence. However, our discussion regarding potential reasons is not derived from this evidence, but rather from our knowledge about software development and machine learning. Consequently, we may overestimate or underestimate and of the issues we discuss as possible reasons in Section~\ref{sec:discussion} and also miss other reasons. However, we clearly state this limitation to be transparent to readers of this article. The only feasible alternative would be not discuss the reasons for our results at all, which we believe would be a larger problem than possible wrong speculative reasoning, if this is clearly marked as such. 

\subsection{External Validity}

While we consider four machine learning libraries in our study, it is unclear if the results would hold for other libraries as well or if a larger effectiveness with more feasible combinations would be possible. However, we do not believe that this is likely, because we already considered large and mature libraries. Additionally, caret is a meta framework that means we actually included many different R libraries in our study. 

We also cannot be certain how our results generalize to other types of machine learning, e.g., clustering or regression. However, we note that these algorithms are similar in their structure to classification algorithms. For example, we see no reason why there should be notable differences in the testability between a random forest regression and a random forest classifier or between a $k$-Nearest Neighbor classifier and a $k$-Means clustering. Thus, we believe that our results provide a strong indication for the expectations beyond classification, i.e., that exactly the same results usually cannot be expected but a more coarse comparison could be possible. 

\subsection{Reliability}

A significant amount of \rev{our} work was the manual analysis of the APIs of four machine learning frameworks we studied carried out together by the authors. We cannot rule out the possibility that we missed algorithms (phase 1) or feasible combinations (phase 2). We believe that the threat for the first phase is negligible, because all the API documentation of all framework specifically contains categories for supervised and/or classification algorithms. For the second phase, we read all API docs and all parameters as pair to minimize the likelihood of missing aspects. Moreover, the results of the second phase were validated by implementing the feasible combinations later on, which required us to double check all results. Thus, while we may have missed something, we do not think that it is likely that we missed so many algorithms or feasible combinations that this would have an effect on our conclusions. 

\section{Conclusion}
\label{sec:conclusion}

Within this paper, we evaluate the potential, feasibility, and effectiveness of differential testing for classification algorithms beyond deep learning. While we found that there is a huge potential, especially for popular algorithms, we found that it was already difficult to identify feasible tests such that the API documentation indicated that the behavior of two implementations should be the same. When we then executed the feasible tests, we found so many differences between the results, that we cannot find a signal in the noise, i.e., identify true positive differences that indicate actual bugs in an implementations among all the false positive tests results with differences due to other reasons. However, our results indicate that for experts it may still be possible to use a relatively lenient approach based on significant differences between classification results to determine if there are bugs within a software. However, the amount of false positive seems to be too large to be useful for researchers as automated pseudo-oracle to evaluate the effectiveness of testing approach for such algorithms.

\bibliography{./literature}

\appendix

\section{Additional Results}

\begin{figure}
\centering
\includegraphics[width=\textwidth]{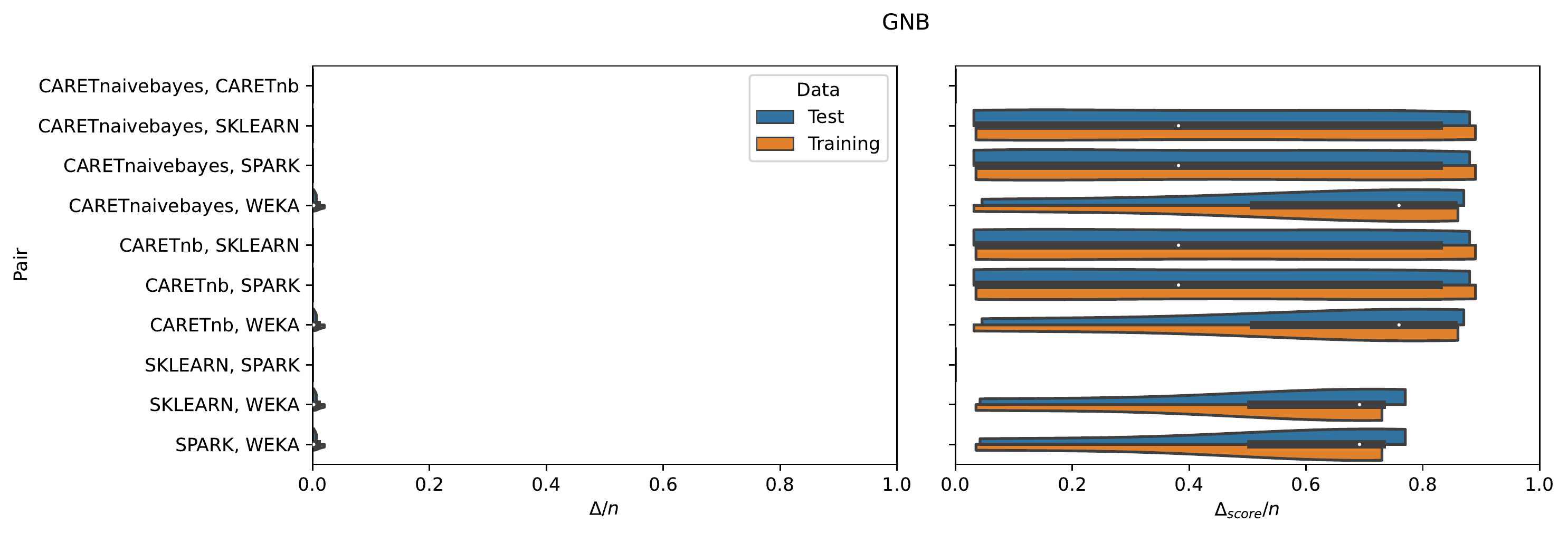}
\caption{\rev{Deviations observed with GNB.}}
\label{fig:results_gnb}
\end{figure}

\begin{figure}
\centering
\includegraphics[width=\textwidth]{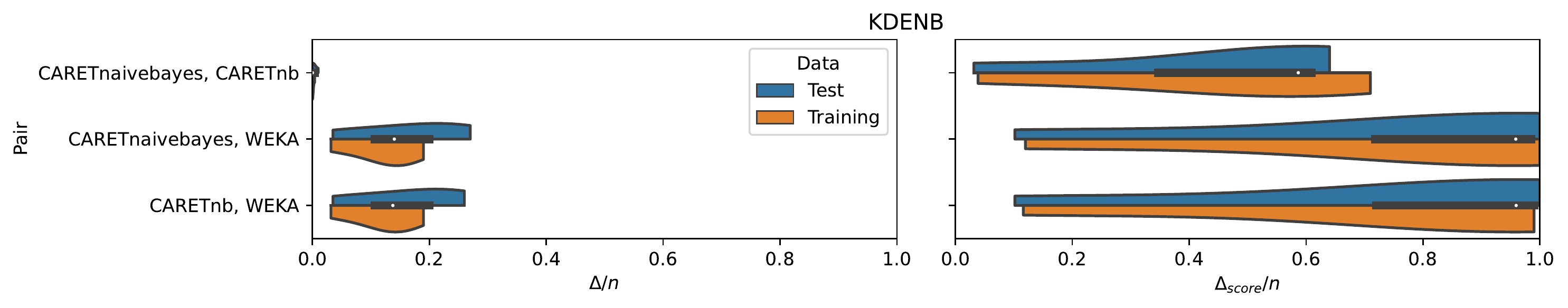}
\caption{\rev{Deviations observed with KDENB.}}
\label{fig:results_kdenb}
\end{figure}

\begin{figure}
\centering
\includegraphics[width=\textwidth]{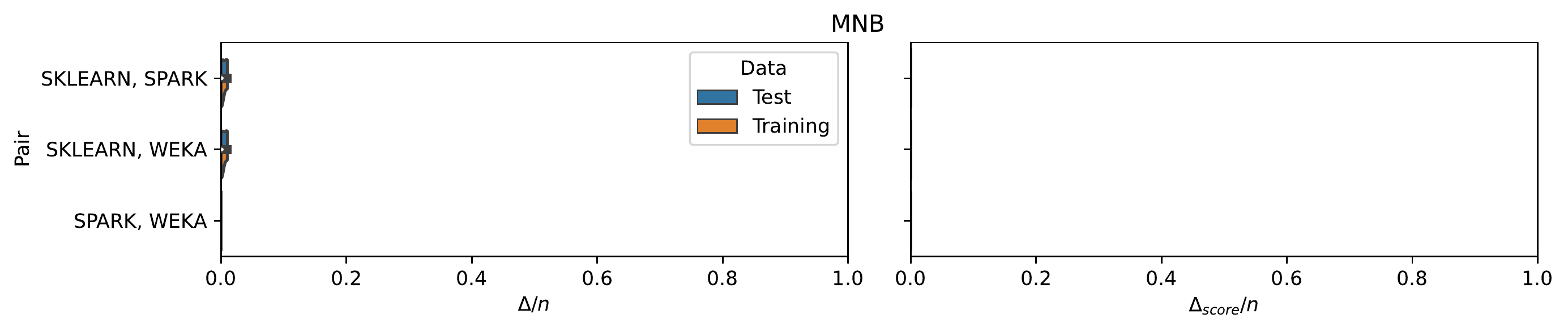}
\caption{\rev{Deviations observed with MNB.}}
\label{fig:results_mnb}
\end{figure}

\begin{figure}
\centering
\includegraphics[width=\textwidth]{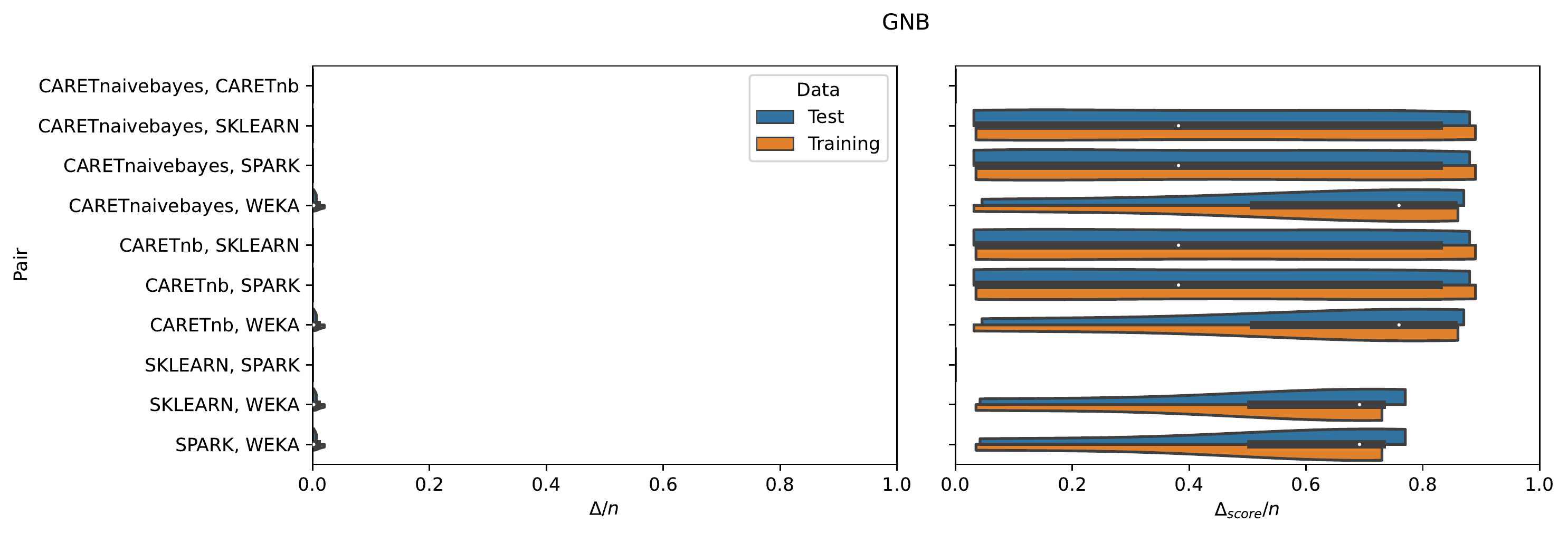}
\caption{\rev{Deviations observed with RF1.}}
\label{fig:results_RF1}
\end{figure}

\begin{figure}
\centering
\includegraphics[width=\textwidth]{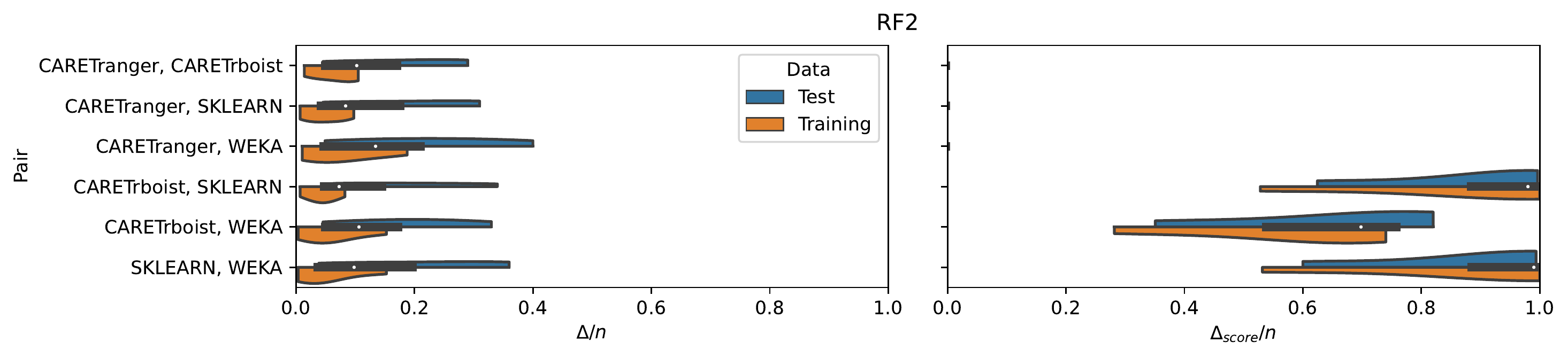}
\caption{\rev{Deviations observed with RF2.}}
\label{fig:results_RF2}
\end{figure}

\begin{figure}
\centering
\includegraphics[width=0.7\textwidth]{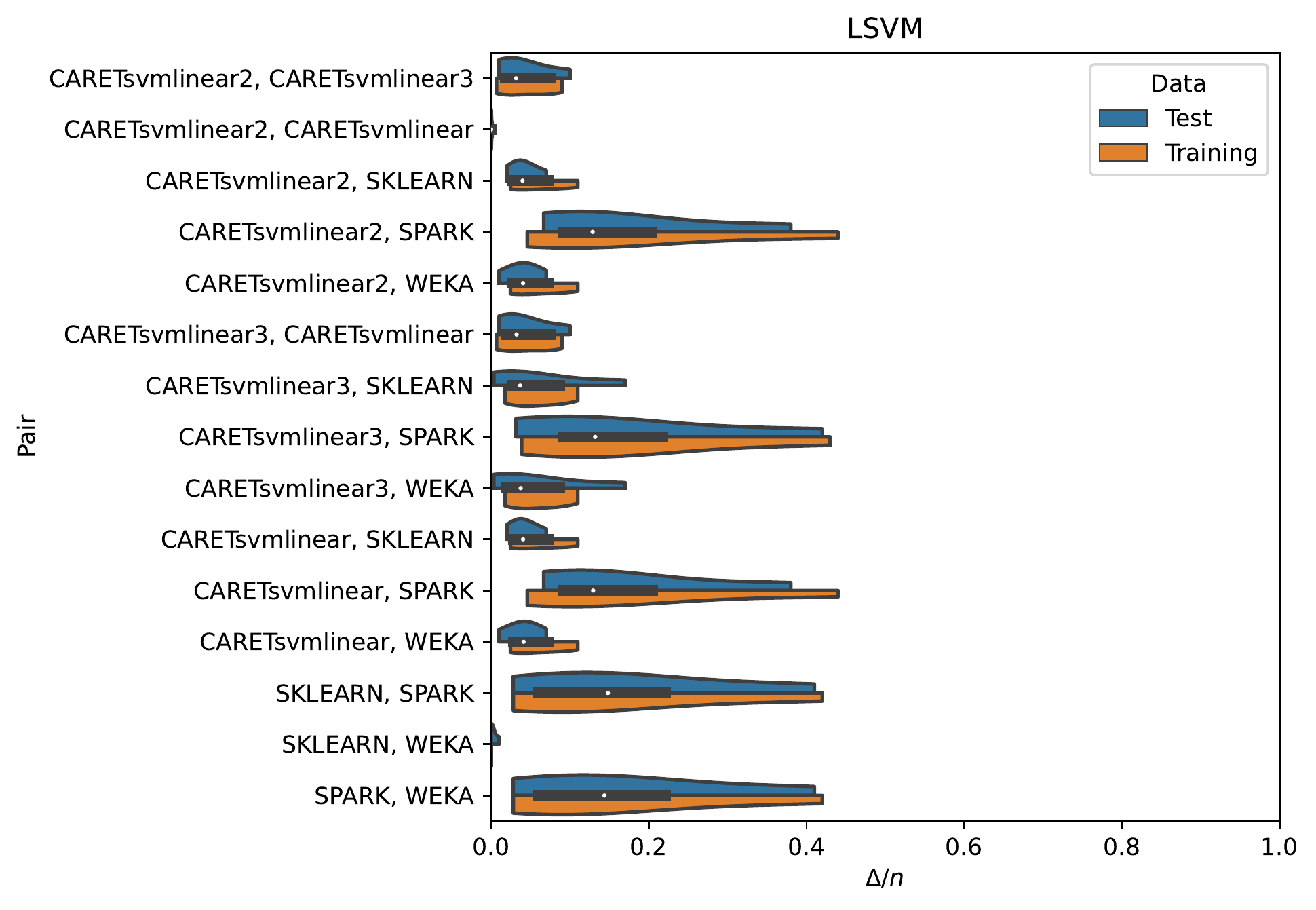}
\caption{\rev{Deviations observed with LSVM.}}
\label{fig:results_lsvm}
\end{figure}

\begin{figure}
\centering
\includegraphics[width=0.7\textwidth]{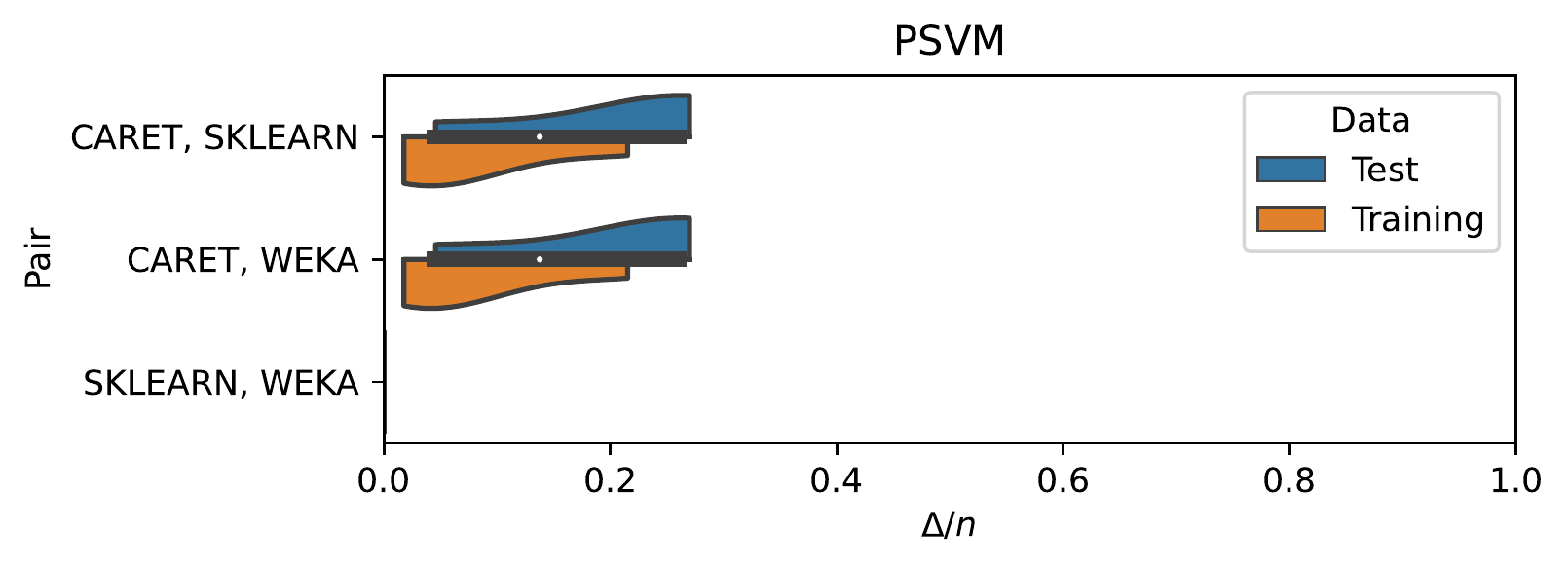}
\caption{\rev{Deviations observed with PSVM.}}
\label{fig:results_psvm}
\end{figure}

\begin{figure}
\centering
\includegraphics[width=0.7\textwidth]{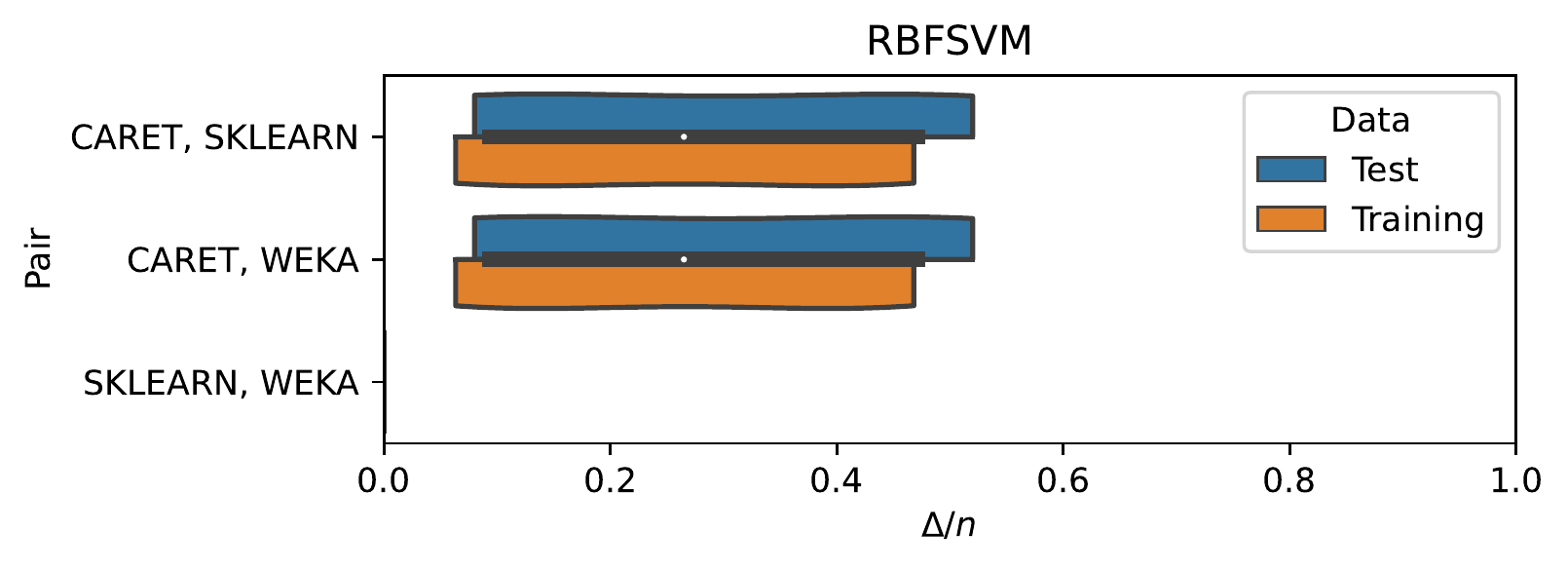}
\caption{\rev{Deviations observed with RBFSVM.}}
\label{fig:results_rbfsvm}
\end{figure}

\begin{figure}
\centering
\includegraphics[width=\textwidth]{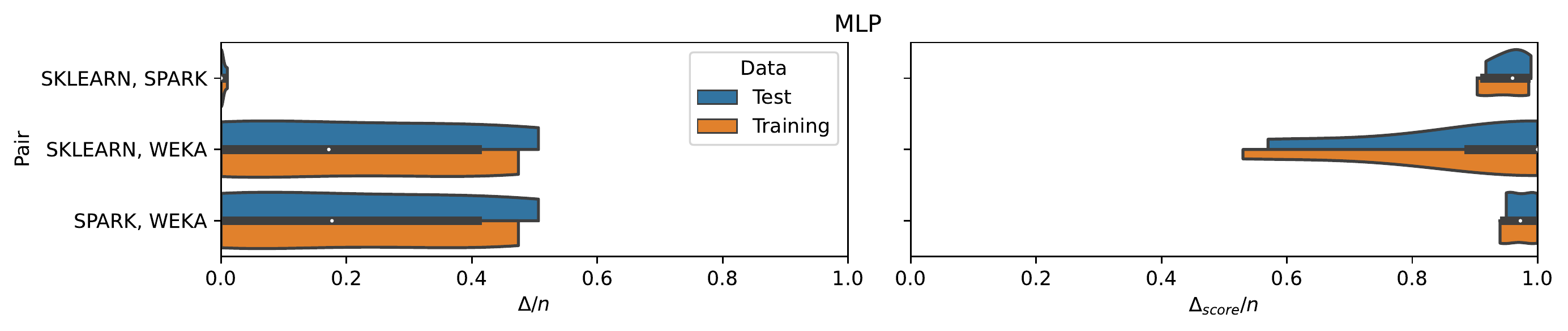}
\caption{\rev{Deviations observed with MLP.}}
\label{fig:results_mlp}
\end{figure}

\begin{figure}
\centering
\includegraphics[width=0.7\textwidth]{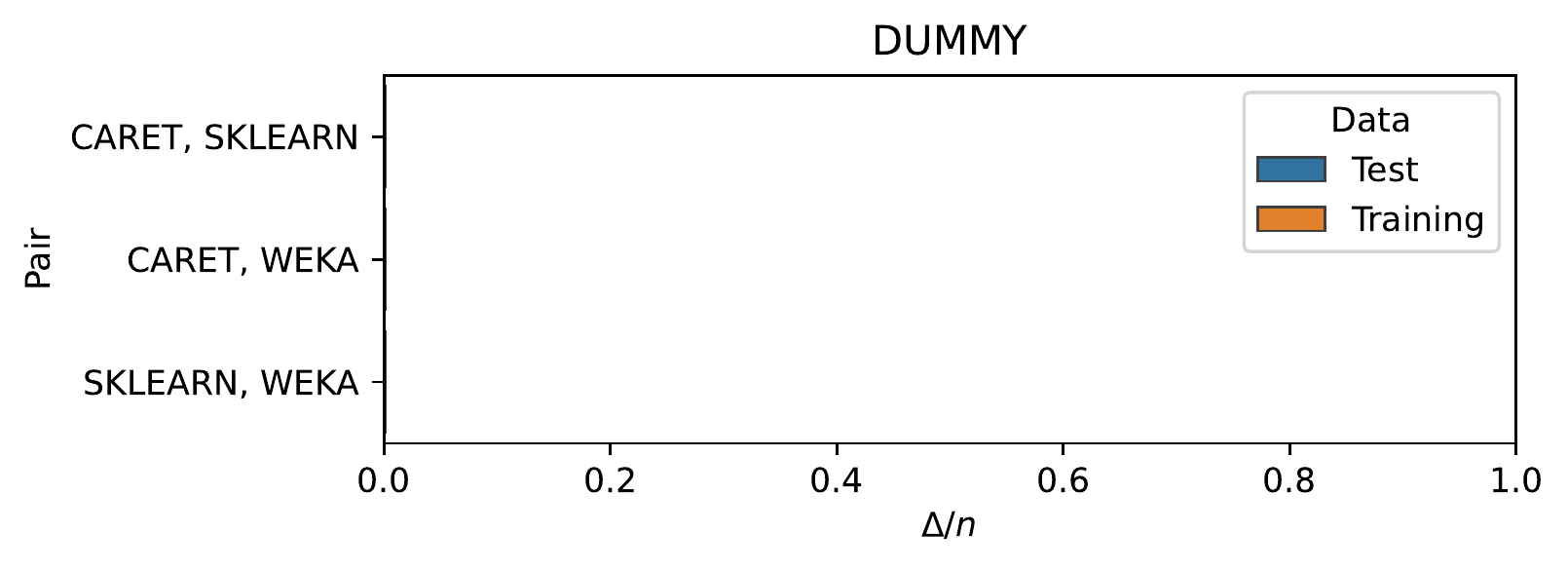}
\caption{\rev{Deviations observed with DUMMY.}}
\label{fig:results_dummy}
\end{figure}

\begin{figure}
\centering
\includegraphics[width=\textwidth]{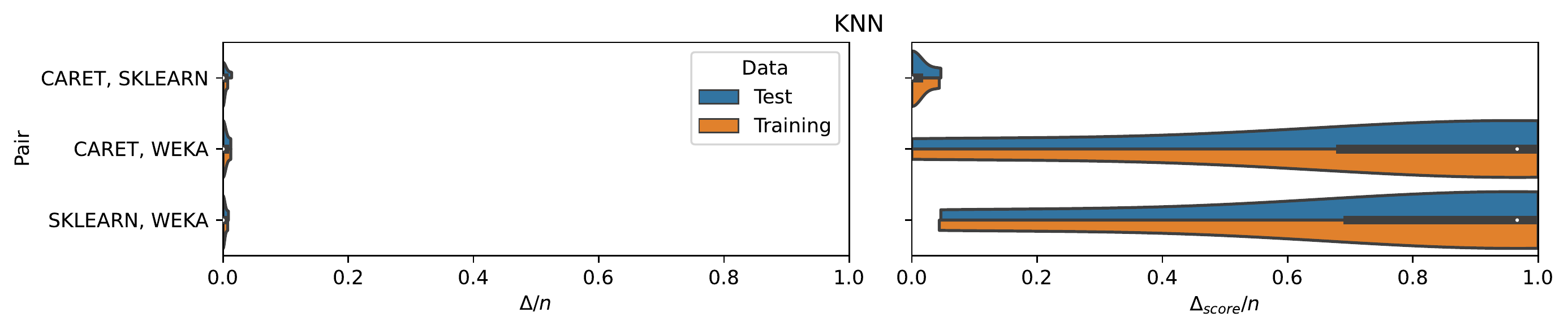}
\caption{\rev{Deviations observed with KNN.}}
\label{fig:results_KNN}
\end{figure}

\begin{figure}
\centering
\includegraphics[width=\textwidth]{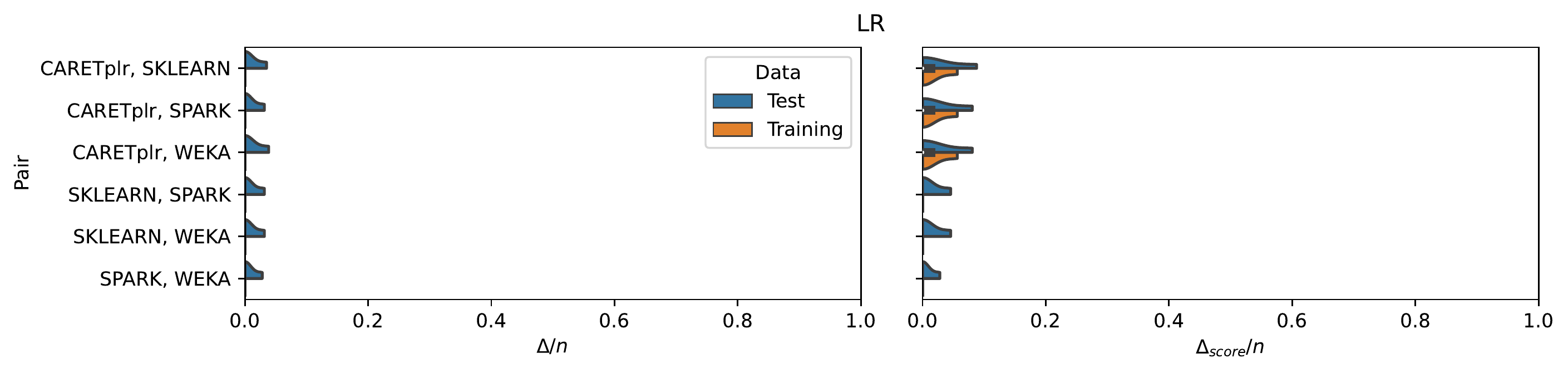}
\caption{\rev{Deviations observed with LR.}}
\label{fig:results_lr}
\end{figure}

\begin{figure}
\centering
\includegraphics[width=\textwidth]{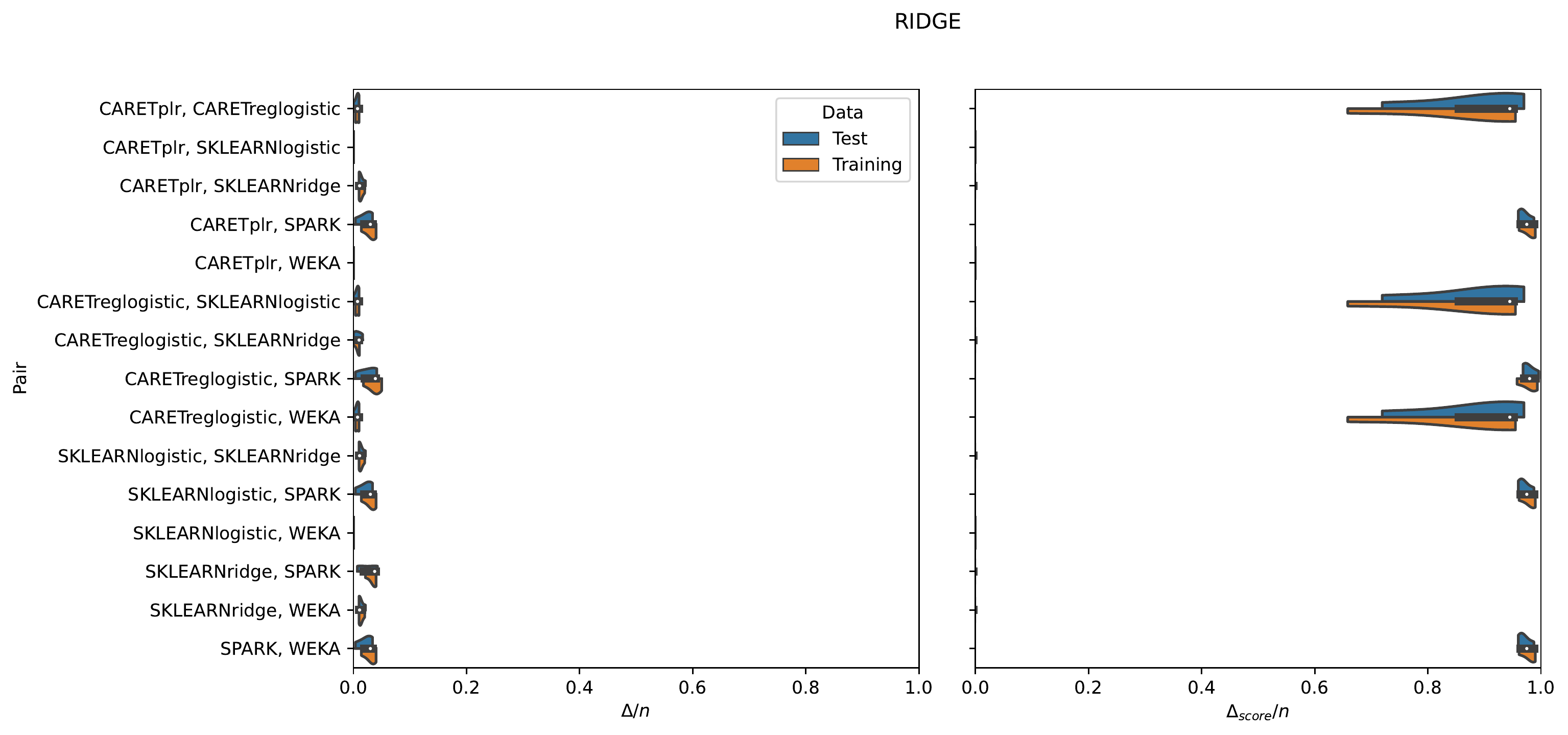}
\caption{\rev{Deviations observed with RIDGE.}}
\label{fig:results_ridge}
\end{figure}

\begin{figure}
\centering
\includegraphics[width=\textwidth]{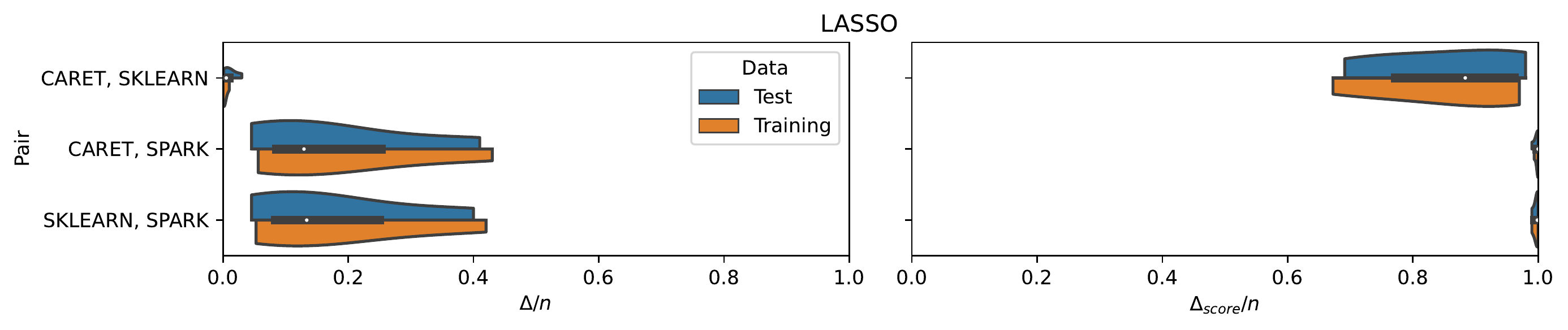}
\caption{\rev{Deviations observed with LASSO.}}
\label{fig:results_lasso}
\end{figure}

Within this appendix, we report additional results for all phases of the case study. Table~\ref{tbl:overlap-full} shows a complete list of algorithms from all frameworks and how they overlap. Table~\ref{tbl:feasible-full} shows how the hyperparameter must be configured for all feasible combinations we identified. Table~\ref{tbl:test-results-full} provides additional details for the results of the execution of the differential tests. \rev{Figure~\ref{fig:results_by_framework} shows how often we observed differences between implementations grouped by the framework they were implemented in. Figure~\ref{fig:results_by_dataset} shows how often we observed differences grouped by the data set we used to execute the tests. Figure~\ref{fig:results_gnb}-\ref{fig:results_lasso} show for each algorithm we tested the distributions of $\Delta$ and $\Delta_{score}$, normalized by the number of instances in the data set $n$.} 

\begin{landscape}
\begin{longtable}{p{4.4cm}p{4.4cm}p{4.4cm}p{4.4cm}}
\caption{Overlapping algorithms between frameworks based on a scan of algorithm names in the API docs.} \label{tbl:overlap-full} \\

\textbf{Scikit-learn} & \textbf{Weka} & \textbf{Spark MLlib} & \textbf{Caret} \\
\hline\hline
\endfirsthead

\multicolumn{4}{c}%
{{\tablename\ \thetable{} -- continued from previous page}} \\
\textbf{Scikit-learn} & \textbf{Weka} & \textbf{Spark MLlib} & \textbf{Caret} \\
\hline\hline
\endhead

\multicolumn{4}{c}{\textit{continued on next page}} \\
\endfoot
\hline
\endlastfoot

BernoulliNB, ComplementNB, MultinomialNB, GaussianNB, CategoricalNB & NaiveBayesMultinomial, NaiveBayes & NaiveBayes & awnb, naive\_bayes, nb, manb
\\\hline
DecisionTreeClassifier & SimpleCart, J48, DecisionStump, HoeffdingTree & DecisionTreeClassifier & C5.0Tree, C5.0Rules, rpartCost, C5.0Cost, rpartScore, rpart, rpart1SE, rpart2, C5.0, ctree, ctree2, bartMachine
\\\hline
RandomForestClassifier & RandomForest & RandomForestClassifier & wsrf, RRF, RRFglobal, rfRules, ordinalRF, ranger, Rborist, rf, ORFlog, ORFpls, ORFridge, ORFsvm, cforest
\\\hline
LinearSVC, NuSVC, SVC & SMO & LinearSVC & svmSpectrumString, svmRadial, svmRadialCost, svmRadialSigma, svmPoly, svmLinear, svmLinear2, svmExpoString, svmRadialWeights, svmBoundrangeString, svmLinearWeights, lssvmRadial, lssvmPoly, lssvmLinear, svmLinear3, svmLinearWeights2
\\\hline
MLPClassifier & MultilayerPerceptron & MultilayerPerceptronClassifier & mlpKerasDecayCost, mlpKerasDecay, mlpKerasDropoutCost, mlpKerasDropout, mlpSGD, mlpML, mlpWeightDecayML, mlp, mlpWeightDecay, monmlp
\\\hline
LogisticRegression, RidgeClassifier & Logistic, SimpleLogistic & LogisticRegression & regLogistic, plr, polr
\\\hline
DummyClassifier & ZeroR &  & null
\\\hline
KNeighborsClassifier & IBk &  & snn, kknn, knn
\\\hline
GradientBoostingClassifier &  & GBTClassifier & gbm\_h2o, gbm
\\\hline
Perceptron & VotedPerceptron &  & 
\\\hline
SGDClassifier & SGD &  & 
\\\hline
GaussianProcessClassifier &  &  & gaussprRadial, gaussprPoly, gaussprLinear
\\\hline
LinearDiscriminantAnalysis &  &  & slda, sparseLDA, sda, rrlda, Linda, rlda, rda, PenalizedLDA, pda, pda2, Mlda, loclda, stepLDA, lda, lda2, RFlda, dda
\\\hline
NearestCentroid &  &  & pam
\\\hline
QuadraticDistriminantAnalysis &  &  & qda, QdaCov, stepQDA
\\\hline
ExtraTreeClassifier &  &  & extraTrees
\\\hline
 & LMT &  & LMT
\\\hline
PassiveAgressiveClassifier &  &  & 
\\\hline
RadiusNeighborsClassifier &  &  & 
\\\hline
 & RandomTree &  & 
\\\hline
 & REPTree &  & 
\\\hline
 & DecisionTable &  & 
\\\hline
 & Ripper &  & 
\\\hline
 & OneR &  & 
\\\hline
 & PART &  & 
\\\hline
 & KStar &  & 
\\\hline
 & BayesNet &  & nbDiscrete
\\\hline
 &  &  & spls, kernelpls, pls, simpls, widekernelpls
\\\hline
 &  &  & gpls, plsRglm
\\\hline
 &  &  & dnn, pcaNNet, mxnet, mxnetAdam, nnet, avNNet, multinom
\\\hline
 &  &  & hda, binda, fda, hdda, hdrda, mda, smda, rmda, amdai
\\\hline
 &  &  & sdwd, dwdLinear, dwdRadial, dwdPoly
\\\hline
 &  &  & xgbDART, xgbLinear, xgbTree
\\\hline
 &  &  & elm
\\\hline
 &  &  & deepboost
\\\hline
 &  &  & glm, glmStepAIC, randomGLM, bayesglm
\\\hline
 &  &  & gam, bam, gamSpline, gamLoess
\\\hline
 &  &  & glmnet, glmnet\_h2o
\\\hline
 &  &  & FRBCS.W, SLAVE, FH.GBML, FRBCS.CHI
\\\hline
 &  &  & rotationForest, rotationForestCp
\\\hline
 &  &  & rbf, rbfDDA
\\\hline
 &  &  & earth, gcvEarth
\\\hline
 &  &  & CSimca, RSimca
\\\hline
 &  &  & tan, tanSearch, awtan
\\\hline
 &  &  & vbmpRadial
\\\hline
 &  &  & ownn
\\\hline
 &  &  & PRIM
\\\hline
 &  &  & nodeHarvest
\\\hline
 &  &  & evtree
\\\hline
 &  &  & nbSearch
\\\hline
 &  &  & xyf
\\\hline
 &  &  & rocc
\\\hline
 &  &  & rFerns
\\\hline
 &  &  & ordinalNet
\\\hline
 &  &  & partDSA
\\\hline
 &  &  & msaenet
\\\hline
 &  &  & lvq
\\\hline
 &  &  & protoclass
\\\hline
 &  &  & vglmCumulative
\\\hline
 &  &  & vglmContRatio
\\\hline
 &  &  & vglmAdjCat
\\\hline
 &  &  & chaid
\\\hline
 &  &  & logreg
\end{longtable}
\end{landscape}

\begin{landscape}
%\begin{longtable}{lp{4cm}p{4cm}p{4cm}p{4cm}}
\begin{longtable}{lllll}
\caption{Feasible combinations with at least three implementations. Algorithm names are italic followed by the required parameters. Parameters in the same row have the same meaning. If parameters are not specified for an algorithm, this means that the API documentation indicates that the default is the same as what is achieved by using the parameters for the other algorithms. The parameter values $x_i$ indicate that the parameter is configurable with the same meaning in all implementations.} \label{tbl:feasible-full} \\

\textbf{Group} & \textbf{Scikit-learn} & \textbf{Weka} & \textbf{Spark MLlib} & \textbf{Caret} \\
\hline\hline
\endfirsthead

\multicolumn{5}{c}%
{{\tablename\ \thetable{} -- continued from previous page}} \\
\textbf{Group} & \textbf{Scikit-learn} & \textbf{Weka} & \textbf{Spark MLlib} & \textbf{Caret} \\
\hline\hline
\endhead

\multicolumn{5}{c}{\textit{continued on next page}} \\
\endfoot
\hline
\multicolumn{5}{p{16cm}}{$^*$Weka SVM kernels are located in the package weka.classifiers.functions.supportVector. The Caret mlp classifiers could not be executed because they require the FCNN4R package, which is not available on CRAN anymore.}
\endlastfoot

GNB & \textit{GaussianNB} & \textit{NaiveBayes} & \textit{NaiveBayes} & \textit{naive\_bayes} \\
& & & modelType=gaussian & usekernel=FALSE \\
& & & & adjust=1 \\
& & & & laplace=0 \\[5pt]
& & & & \textit{nb} \\
& & & & usekernel=FALSE \\
& & & & adjust=1 \\
& & & & laplace=0 \\
\hline
KDENB & & \textit{NaiveBayes} & & \textit{naive\_bayes} \\
& & -K & & usekernel=TRUE \\
& & & & adjust=1 \\
& & & & laplace=0 \\[5pt]
& & & & \textit{nb} \\
& & & & usekernel=TRUE \\
& & & & adjust=1 \\
& & & & laplace=0 \\
\hline
MNB & \textit{MultinomialNB} & \textit{NaiveBayesMultinomial} & \textit{NaiveBayes} & \\
& & & modelType=multinomial & \\
\hline
RF1 & \textit{RandomForestClassifier} & \textit{RandomForest} & & \textit{ranger} \\
& n\_estimators=$x_1$ & -I $x_1$ & & num.trees=$x_1$ \\
& max\_features=$x_2$ & -K $x_2$ & & mtry=$x_2$ \\
& max\_depth=$x_3$ & -depth $x_3$ & & max.depth=$x_3$ \\
& & & & splitrule=gini \\
& & & & min.node.size=1 \\[5pt]
& & & & \textit{rborist} \\
& & & & nTree=$x_1$ \\
& & & & predFixed=$x_2$ \\
& & & & nLevel=$x_3$ \\
& & & & minNode=1 \\
\hline
RF2 & \textit{RandomForestClassifier} & \textit{RandomForest} & & \textit{ranger} \\
& n\_estimators=$x_1$ & -I $x_1$ & & num.trees=$x_1$ \\
& max\_features=$x_2$ & -K $x_2$ & & mtry=$x_2$ \\
& & & & splitrule=gini \\
& & & & min.node.size=1 \\[5pt]
& & & & \textit{rborist} \\
& & & & nTree=$x_1$ \\
& & & & predFixed=$x_2$ \\
& & & & minNode=1 \\[5pt]
& & & & \textit{rf} \\
& & & & ntree=$x_1$ \\
& & & & mtry=$x_2$ \\
\hline
LSVM & \textit{SVC} & \textit{SMO} & \textit{LinearSVC} & \textit{svmLinear} \\
& C=$x_1$ & C $x_1$ & regParam=$x_1$ & tau=$x_1$ \\
& kernel=linear & -k PolyKernel$^*$ & \\
& & & & tol=0.001 \\[5pt]
& & & & \textit{svmLinear2} \\
& & & & cost=$x_1$ \\[5pt]
& & & & \textit{svmLinear3} \\
& & & & cost=$x_1$ \\
& & & & epsilon=0.001 \\
& & & & Loss=L2 \\
\hline
PSVM & \textit{SVC} & \textit{SMO} & & \textit{svmPoly} \\
& C=$x_1$ & C $x_1$ & & C=$x_1$ \\
& degree=$x_2$ & -E $x_2$ & & degree=$x_2$ \\
& kernel=poly & -k PolyKernel$^*$ & & \\
& gamma=1 & & & scale=1 \\
& & & & tol=0.001 \\
\hline
RBFSVM & \textit{SVC} & \textit{SMO} & & \textit{svmRadial} \\
& C=$x_1$ & C $x_1$ & & C=$x_1$ \\
& gamma=$x_2$ & -G $x_2$ & & sigma=$x_2$ \\
& kernel=rbf & -k RBFKernel$^*$ & & \\
& & & & tol=0.001 \\
\hline
MLP & \textit{MLPClassifier} & \textit{MultilayerPerceptron} & \textit{MultilayerPerceptronClassifier} & \textit{mlp}* \\
& hidden\_layer\_sizes=$x_1$ & -h $x_1$ & layers=$x_1$ & size=$x_1$ \\
& learning\_rate\_init=$x_2$ & -L $x_2$ & stepSize=$x_2$ & learnFuncParams=c($x_1$,0) \\
& max\_iter=$x_3$ & -N $x_3$ & maxIter=$x_3$ & maxit=$x_3$ \\
& momentum=0.0 & -M 0 & & \\
& solver=sgd & & solver=gd & \\
& alpha=0.0 & & & \\
& activation=logistic & & & \\
& & -I & & \\[5pt]
& & & & \textit{mlpSGD}* \\
& & & & size=$x_1$ \\
& & & & learn\_rate=$x_2$ \\
& & & & max\_epochs=$x_3$ \\
& & & & momentum=0 \\
& & & & l2reg=0 \\
& & & & gamma=0 \\
& & & & lambda=0 \\
& & & & repeats=1 \\
\hline
DUMMY & \textit{DummyClassifier} & \textit{ZeroR} & & \textit{null} \\
& strategy=most\_frequent & & & \\
\hline
KNN & \textit{KNeighborsClassifier} & \textit{IBk} & & \textit{knn} \\
& n\_neighbors=$x_1$ & -K $x_1$ & & k=$x_1$ \\
\hline
LR & \textit{LogisticRegression} & \textit{Logistic} & \textit{LogisticRegression} & \textit{regLogistic} \\
& & -R 0 & regParam=0 & cost=0 \\
& max\_iter=10000 & -M 10000 & maxIter=10000 & \\
& & -S & & \\
& penalty=none & & & loss=L2\_primal \\
& & & & epsilon=0.0001 \\[5pt]
& & & & \textit{plr} \\
& & & & lambda=0 \\
& & & & cp=bic \\[5pt]
& & & & \textit{polr} \\
& & & & method=logistic \\
\hline
RIDGE & \textit{LogisticRegression} & \textit{Logistic} & \textit{LogisticRegression} & \textit{regLogistic} \\
& C=$1/(2\cdot x_1)$ & -R $x_1$ & regParam=$x_1$ & cost=$x_1$ \\
& max\_iter=10000 & -M 10000 & maxIter=10000 & \\
& & -S & & \\
& penalty=l2 & & & loss=L2\_primal \\
& & & & epsilon=0.0001 \\[5pt]
& \textit{RidgeClassifier} & & & \textit{plr} \\
& alpha=$x_1$ & & & lambda=$x_1$ \\
& max\_iter=10000 & & & \\
& & & & cp=bic \\
\hline
LASSO & \textit{LogisticRegression} & & \textit{LogisticRegression} & \textit{regLogistic} \\
& C=$1/x_1$ & & regParam=$x_1$ & cost=$x_1$ \\
& max\_iter=10000 & & maxIter=10000 & \\
& penalty=l1 & & elasticNetParam=1 & loss=L1 \\
& & & & epsilon=0.0001 \\
\end{longtable}
\end{landscape}

\renewcommand*{\arraystretch}{1.5}
\begin{longtable}{l p{9.5cm}}
\caption{Feasible combinations with at least three implementations. Algorithm names are italic followed by the required parameters. Parameters in the same row have the same meaning. If parameters are not specified for an algorithm, this means that the API documentation indicates that the default is the same as what is achieved by using the parameters for the other algorithms. The parameter values $x_i$ indicate that the parameter is configurable with the same meaning in all implementations.} \label{tbl:test-results-full} \\

\textbf{Group} & \textbf{Test results} \\
\hline
\endfirsthead

\multicolumn{2}{c}%
{{\tablename\ \thetable{} -- continued from previous page}} \\
\textbf{Group} & \textbf{Test results} \\
\hline\hline
\endhead

\multicolumn{2}{c}{\textit{continued on next page}} \\
\endfoot

\endlastfoot
\textbf{GNB} & The classes of all implementations are equal, except Weka, which yields different classes on up to 2\% of the instances on the RANDOM, WINE, and BC data. These differences are not significant. The scores of the two pair naive\_bayes and nb, as well as the pair Scikit-learn Spark MLlib are equal. For all other pairs, we often observe large numbers of deviations between the scores on up to 89\% of the instances, both on the test and training data. However, these differences are also not significant, i.e., we observe the same distributions when all instances are considered, but the scores of the individual instances depend on the implementation.\\
\textbf{KDENB} & We observe differences between all implementations. The differences between the R packages naive\_bayes and nb are small, with respect to $\Delta$, i.e, we observe only few differences in classifications with at most 1\%. However, we observe big differences in scores, on both training and test data. However, none of these differences are significant. The differences to the Weka implementation are large: we never observed the exact same classification with $\Delta$ showing differences on between 3\% and 27\% of the data. These differences are significant in about half of the time, i.e., on the WINE data and the test date of UNIFORM. We also observe large differences in the scores, including cases where every single score is different. Similar to the classifications, the differences in scores are significant one the WINE data, the test data of UNIFORM, and, in addition, the test data of RANDOM, but only between the nb package and WEKA. \\
\textbf{MNB} & We did not observe any differences between the Weka and Spark MLlib implementation. We also did not observe any differences in the scores between all three implementations. However, the Scikit-learn implementation classifies one instance differently in  ten of 21 comparisons. These differences are in cases where the score is almost exactly 0.5 and for one framework slightly smaller and for the other framework slightly larger than 0.5, e.g., for Scikit-learn 0.499 and for Weka 0.501. In this case, we do not consider the scores different, but the classes are different regardless. \\
\textbf{RF1} & The results of the random forests with a fixed depth of five are never exactly the same. Between 0.7\% and 40\% of classifications are different. We note that the differences are larger on the test data than on the training data. The differences between the two R packages and Scikit-learn are similar to each other. The deviations between these three implementations and Weka are about twice as large. However, these differences between the classifications are mostly not significant, with two exceptions on the RANDOM data, where ranger deviates significantly from Scikit-learn and weka. For the scores, we observe that the Weka and Rborist have deviations on between 28\% and 82\% of the instances. However, these large differences are only significant on the training data of WINE. The scores of Weka imlementation and Rborist have deviations from Scikit-learn on between 52\% and 100\% of the instances. These differences are almost always significant. We note that package ranger does not support scores. \\
\textbf{RF2} & The results of the random forest without a fixed depth are also almost never exactly the same, with two exceptions, once for Scikit-learn and the R package ranger, and once for Scikit-learn and the R package rf. Otherwise, there are differences for up to 38\% of the instances. Same as for the random forest with fixed depth, the differences are larger on the test data. However, in most cases, these differences are not significant, except on the WINE data, where ranger is significantly different from all other implementations. For the scores differences on between 23\% and 87\% of the instances. However, none of these differences are significant, i.e., while the scores are different, their distributions are not.\\
\textbf{LSVM} & None of the linear SVMs implementations lead to equal results. The svmlinear and svmlinear2 from Caret are almost equal, with only a single instance that is classified differently on the WINE data. The Scikit-learn and Weka implementations are also almost equal with two single instance misclassified on the test data of UNIFORM and WINE. The differences between most other pairs of implementations are similar, with between 1\% and 11\% of the instances deviating from each other. The clear outlier is the Spark MLlib implementation, which yields a different class on 38\% to 44\% of the instances on the RANDOM data, and also differences on 12\% to 16\% of the WINE instances. These outliers of Spark MLlib are the only significant deviations of the linear SVM. \\
\textbf{PSVM} & The Scikit-learn and Weka implementations of the polynomial SVM are equal. The Caret implementation yields different classifications for 17\% to 27\% of the instances. However, this difference is only significant on the UNIFORM data. \\
\textbf{RBFSVM} & The Scikit-learn and Weka implementations of the RBF SVM are equal. The care implementation yields different classifications for 6\% to 52\% of the instances. This difference is always significant, except training data of BC.\\
\textbf{MLP} & All three MLPs lead to the same classifications on the RANDOM data. The Scikit-learn and Weka implementations are also also equal on the UNIFORM data. The Scikit-learn and Spark MLlib implementation classify one instance differently on the UNIFORM data. However, they are equal on the BC and WINE data. The difference of Scikit-learn and Spark MLlib to Weka is on the BC and WINE data is significant, with differences on between 34\% and 51\% of the instances. The scores of all three implementations are significantly different from each other, with deviations on between 53\% and 100\% of the instances. \\
\textbf{DUMMY} & The trivial classifiers always have the same classes. The scores depend on the implementation of the trivial model: Caret and Weka have the same approach and always yield the same results. Scikit-learn uses a different approach and disagrees with the two other scores. These disagreements are significant. \\
\textbf{KNN} & The nearest neighbor algorithms have the same classes, except on the WINE data, where about 1\% of the instances are classified differently between all combinations of frameworks. These differences are not significant. The scores of Scikit-learn and Caret are also equal, except on the WINE data, where 5\% of the instances have different scores. This difference is significant. On WINE, the scores of Caret are equal to those of Weka. On the other data sets, Weka has significantly different scores from Caret and Scikit-learn, with 91\% to 100\% of instances receiving different scores. \\
\textbf{LR} & The logistic regressions have the same classes, except on the test data of BC, where about 3\% of the instances are classified differently. These differences are not significant. Similarly, the scores are always equal for all implementations  on the UNIFORM, RANDOM, and WINE data, as well as for Scikit-learn, Weka, and Spark MLlib on the BC training data. We observe differences on between 3\% and 9\% of the instances on the remaining tests on the BC data. The differences between the scores are significant. \\
\textbf{RIDGE} & The Weka, the LogisticRegression from Scikit-learn, and the Caret model plr yield identical for both classifications and scores results. The differences between the classes predicted by the other pairs of implementations are between 0.3\% and 4\%, i.e., relatively small and not statistically significant. The scores are inconsistent and have large and significant deviations of on between 65\% and 99\% of the instances. The RidgeClassifier of Scikit-learn does not compute scores. \\
\textbf{LASSO} & The Caret and Scikit-learn implementation yield almost the same classes, with up to three instances classified differently. These differences are not significant. The differences to Spark MLlib are large and between 4\% and 43\% of instances are classified differently. However, these differenes are only significant on the RANDOM data. The scores are different for between 67\% and 98\% of the instances for Caret and Scikit-learn. However, the difference is only significant on the training data of BC. The scores of Spark MLlib are different on at least 99\% and 100\% of the instance, i.e., almost always. These differences are significant. 
\end{longtable}

\end{document}